\documentclass[10pt,a4paper]{iopart}
\usepackage{epsf}

\let\idxI=\kappa \let\idxII=\lambda \let\idxIII=\mu    
\def\Sch{Schwarzschild}                             
\def\be{\begin{equation}} \def\ee{\end{equation}}   
\def\bea{\begin{eqnarray}} \def\eea{\end{eqnarray}} 
\def\d{{\rm d}}                                
\def\e{{\rm e}}                                
\def\expon#1{\e^{#1}}                           
\def\oder#1#2{\frac{\d #1}{\d #2}}               
\let\bra=\langle                                             
\def\nd{\discretionary{--}{--}{--}}   
\begin{document}
\jl{6}
\title[Optical reference geometry of the Kerr\nd Newman spacetimes]%
  {Optical reference geometry of the Kerr\nd Newman spacetimes%
  \footnote{Published in Class.\ Quantum Grav.\ \textbf{17} (2000),
    pp.~2691--2718.}}
\author{Z Stuchl\'{\i}k\footnote{E-mail address: Zdenek.Stuchlik@fpf.slu.cz},
  S Hled\'{\i}k\footnote{E-mail address: Stanislav.Hledik@fpf.slu.cz}
  and J. Jur\'a\v{n}\footnote{E-mail address: Josef.Juran@fpf.slu.cz}}
\address{Institute of Physics, Faculty of Philosophy and Science,
  Silesian University in~Opava, Bezru\v{c}ovo n\'am.~13,
  CZ-746\,01~Opava, Czech Republic}

\begin{abstract}
  Properties of the optical reference geometry related to Kerr\nd New\-man
  black-hole and naked-singularity spacetimes are illustrated using
  embedding diagrams of their equatorial plane. It is shown that among all
  inertial forces defined in the framework of the optical geometry, just
  the centrifugal force plays a fundamental role in connection to the
  embedding diagrams because it changes sign at the turning points of the
  diagrams.  The embedding diagrams do not cover the stationary part of the
  Kerr\nd Newman spacetimes completely. Hence, the limits of embeddability
  are given, and it is established which of the photon circular orbits
  hosted the by Kerr\nd Newman spacetimes appear in the embeddable regions.
  Some typical embedding diagrams are constructed, and the Kerr\nd Newman
  backgrounds are classified according to the number of embeddable regions
  of the optical geometry as well as the number of their turning points.
  It is shown that embedding diagrams are closely related to the notion of
  the radius of gyration which is useful for analyzing fluid rotating in
  strong gravitational fields.
\end{abstract}


\maketitle

\section{Introduction}

The optical reference geometry related to stationary spacetimes enables to
introduce the concept of inertial forces in the framework of general
relativity in a natural way \cite{ACL,ANW93}. Of course, in accord with the
spirit of general relativity, alternative approaches to the concept of
inertial forces are possible (see, e.g., \cite{Sem,JCB}); however, here we
shall follow the approach of Abramowicz and his coworkers \cite{ANW95},
providing a description of relativistic dynamics in accord with Newtonian
intuition.

The optical geometry results from an appropriate conformal ($3+1$)
splitting, reflecting some hidden properties of the spacetimes under
consideration through its geodesic structure.  The geodesics of the optical
geometry related to static spacetimes coincide with trajectories of light,
thus being `optically straight' \cite{AP,AMS}.  Moreover, the geodesics are
`dynamically straight,' because test particles moving along them are kept
by a velocity-independent force \cite{A90}; they are also `inertially
straight,' because gyroscopes carried along them do not precess along the
direction of the motion \cite{A92}.

Some properties of the optical geometry can be appropriately demonstrated
by embedding diagrams of its representative sections \cite{ACL,KSA,SH99a}.
Because we are familiar to the Euclidean space, usually 2-dimensional
sections of the optical space are embedded into the 3-dimensional Euclidean
space. (Of course, embeddings into other conveniently chosen spaces can
also provide interesting information, however, we shall focus our attention
on the most straightforward Euclidean case.)  In the Kerr\nd Newman
backgrounds, the most representative section is the equatorial plane, which
is their symmetry plane. This plane is also of great astrophysical
importance, especially in connection to the theory of accretion disks
\cite{B}.

In  the  spherically  symmetric spacetimes  (Schwarzschild  \cite{AP},
Reissner\nd  Nord\-str\"om \cite{KSA}, and  Schwarzschild\nd de~Sitter
\cite{SH99b}), an interesting  coincidence appears: the turning points
of the central-plane  embedding diagrams of the optical  space and the
photon circular orbits are located at the same radii, where, moreover,
the  centrifugal force,  related to  the optical  space,  vanishes and
reverses sign.

However, in the rotating black-hole and naked-singularity backgrounds, the
centrifugal force does not vanish at the radii of photon circular orbits in
the equatorial plane \cite{IP}.  Of course, the same statement is true if
these rotating backgrounds carry a nonzero electric charge.  It is,
therefore, interesting to study how the inertial forces, defined in the
framework of the optical geometry, and the photon circular orbits are
related to equatorial-plane embedding diagrams of the optical geometry of
rotating, charged backgrounds. Such relations were discussed in the case of
non-charged, Kerr backgrounds in \cite{SH99c}. In this paper, we shall
generalise the results to the more complex case of the Kerr\nd Newman
backgrounds, in which even stable photon circular orbits can exist beside
the unstable ones, contrary to the case of Kerr backgrounds \cite{BBS}.

In Section \ref{ogif}, the optical reference geometry and the related
inertial forces are defined relative to the family of locally non-rotating
observers in the Kerr\nd Newman spacetimes. In Section \ref{secm},
stationary equatorial circular motion in the Kerr\nd Newman spacetimes is
discussed, using the concept of the gravitational and inertial forces
expressed in terms of a `Newtonian' velocity related to the optical
geometry. It is shown that asymptotically the relativistic expressions of
the gravitational, Coriolis and centrifugal forces reduce to the well known
Newtonian formulae. In central parts, they enable an illumination of
unusual properties of the Kerr\nd Newman spacetimes in terms of intuitively
clear concepts. The properties of the centrifugal force in the equatorial
plane are closely related to the properties of the embedding diagrams of
the optical-geometry equatorial plane. In Section \ref{edog}, the embedding
formula is introduced, the limits of Euclidean embeddability of the optical
geometry are established, and the turning points of the embedding diagrams
are determined; it is also shown that they occur just where the centrifugal
force reverses sign.  The locations of photon circular orbits in the
equatorial plane are given, and it is established which of them are
contained in the embeddable regions of the optical space.  The Kerr\nd
Newman spacetimes are classified according to the criterion of
embeddability of regions containing photon circular orbits.  Finally,
typical embedding diagrams are constructed, and the Kerr\nd Newman
spacetimes are classified according to the properties of the embedding
diagrams (namely, the numbers of embeddable regions and turning points of
the diagrams).  In Section \ref{concluding}, some concluding remarks are
presented.

\section{Optical geometry and inertial forces}            \label{ogif}

The notions of the optical reference geometry and related inertial forces
are convenient for spacetimes with symmetries, especially for stationary
(static) and axisymmetric (spherically symmetric) ones. However, they can
be introduced for a general spacetime lacking any symmetry \cite{ANW93}.

\subsection{General case}

Assuming a hypersurface globally  orthogonal to a timelike unit vector
field $n^\idxI$ and a scalar function $\Phi$ satisfying the conditions
\be
  n_{[\idxI} \nabla\!_\idxII n_{\idxIII]} = 0,                   \quad
  n^\idxI n_\idxI = -1,                                          \quad
  \dot{n}_\idxII = n^\idxI \nabla\!_\idxI n_\idxII =
    \nabla\!_\idxII \Phi,
\ee
the 4-velocity  $u^\idxI$ of a test  particle of rest mass  $m$ can be
uniquely decomposed as
\be                                                     
  u^\idxI = \gamma (n^\idxI + v\tau^\idxI).                 \label{e7}
\ee
Here $\tau^\idxI$ is a unit vector orthogonal to $n^\idxI$, $v$ is the
speed and $\gamma = (1-v^2)^{-1/2}$.

Introducing, according to Abramowicz, Nurowski and Wex \cite{ANW93}, a
projected  3\discretionary{-}{-}{-}space orthogonal to  $n^\idxI$ with
the  positive  definite metric  giving  so  called ordinary  projected
geometry
\be                                          
  h_{\idxI\idxII} = g_{\idxI\idxII} + n_\idxI  n_\idxII,
\ee
and  the  optical   geometry  $\tilde{h}_{\idxI\idxII}$  by  conformal
rescaling
\be
  \tilde{h}_{\idxI\idxII} =
    \expon{-2\Phi} (g_{\idxI\idxII} + n_\idxI n_\idxII),    \label{e9}
\ee
the projection of  the 4-acceleration $a^\bot_\idxI = h^\idxII_\idxI\,
u^\idxIII \nabla\!_\idxIII  u_\idxII$ can be  uniquely decomposed into
terms  proportional  to  zeroth,  first  and  second  powers  of  $v$,
respectively, and the velocity change
\be
  \dot{v} = (\expon{\Phi}\gamma v)_{,\idxIII}\,u^\idxIII.
\ee
Thus, we  arrive to covariant definition of  inertial forces analogous
to the Newtonian physics \cite{ANW93,ACNSV}
\be
  ma^\bot_\idxI =
    G_\idxI(v^0) +
    C_\idxI(v^1) +
    Z_\idxI(v^2) +
    E_\idxI(\dot{v}),
\ee
where the first term
\be                                               
  G_\idxI = - m \nabla\!_\idxI \Phi = - m \Phi\!_{,\idxI}   \label{gf}
\ee
corresponds to the gravitational force, the second term
\be                                                
  C_\idxI = - m \gamma^2 v n^\idxII
    (\nabla\!_\idxII \tau_\idxI -
    \nabla\!_\idxI \tau_\idxII)                           \label{cltf}
\ee
corresponds to the Coriolis\nd Lense\nd Thirring force, the third term
\be                                              
  Z_\idxI =
    -m (\gamma v)^2
    \tilde{\tau}^\idxII
    \tilde{\nabla}\!_\idxII \tilde{\tau}_\idxI              \label{cf}
\ee
corresponds to the centrifugal force, and the last term
\be                                              
  E_\idxI = - m \dot{v} \tilde{\tau}_\idxI                  \label{ef}
\ee
corresponds to the Euler  force. Here $\tilde{\tau}^\idxI$ is the unit
vector   along    $\tau^\idxI$   in   the    optical   geometry,   and
$\tilde{\nabla}\!_\idxI$ is  the covariant derivative  with respect to
the optical geometry.

\subsection{Kerr\nd Newman case}

Using geometric system of units  ($c=G=1$), and denoting $M$ the mass,
$a$ the specific  angular momentum, $e$ the electric  charge, the line
element  of  the  Kerr\nd  Newman  spacetime, expressed  in  terms  of
standard Boyer\nd Lindquist coordinates, reads
\bea                                                  
  \d s^2 &=&
    -\left(1- \frac{2Mr-e^2}{\Sigma}\right)\,\d t^2 -
    \frac{2a(2Mr-e^2)}{\Sigma}\sin^2\theta\,\d t\d\phi +  \nonumber \\
    &&\frac{A\sin^2\theta}{\Sigma}\,\d\phi^2 +
    \frac{\Sigma}{\Delta}\,\d r^2 +
    \Sigma\,\d\theta^2,                                     \label{e1}
\eea
where
\bea                                                  
  &&\Delta = r^2 - 2Mr + a^2 + e^2,               \\
  &&\Sigma = r^2 + a^2 \cos^2\theta,              \\
  &&A = (r^2+a^2)^2 - \Delta a^2 \sin^2 \theta.
\eea
For simplicity  we put  $M=1$ in the  following; equivalently,  we use
mass units  of $M$\@.  If $a^2  + e^2 \leq  1$, the  metric (\ref{e1})
represents black-hole  spacetimes. The  loci of their  horizons, $r_-$
(the inner one) and $r_+$ (the outer one), determined by real roots of
$\Delta(r;a,e) = 0$, can be equivalently given by the relation
\be                                                  
  a^2 = a^2_{\rm h}(r;e) = r(2-r)- e^2.
\ee
The case $a^2 + e^2 = 1$ corresponds to extreme black holes. If $a^2 +
e^2 > 1$, there are  no horizons, and the metric (\ref{e1}) represents
a naked-singularity spacetime.  If $e^2  < 1$, both the black-hole and
naked-singularity  spacetimes contain an  ergosphere, where  $g_{tt} <
0$; particles and photons in  bound states with covariant energy $E<0$
are  possible  there  \cite{MTW}.  Naked-singularity  spacetimes  with
$e^2\geq 1$ have no ergosphere.

The Kerr\nd Newman spacetimes, being stationary and axially symmetric,
admit  two   commuting  Killing   vector  fields;  the   vector  field
$\eta^\idxI$  is  (at  least  asymptotically)  timelike,  having  open
trajectories, the vector field $\xi^\idxI$ is spacelike, having closed
trajectories.    Now,  the   vector  field   $n^\idxI$   relevant  for
constructions  of  the ordinary  projected  geometry  and the  optical
reference geometry can  be given by using these  Killing vector fields
\cite{ANW95}, and  corresponds to  the family of  locally non-rotating
frames (LNRF) or zero  angular momentum observers (ZAMO) introduced by
Bardeen \cite{B}. Namely,
\be
\begin{array}{l}         
  n^\idxI =
    \expon{\Phi} (\eta^\idxI + \Omega_{\rm LNRF} \xi^\idxI),     \quad
  \Omega_{\rm LNRF} =
    -\xi^\idxII \eta_\idxII / \eta^\idxIII \eta_\idxIII,            \\
  \Phi = -{\textstyle\frac{1}{2}}
    \ln\left(-\eta^\idxII \eta_\idxII -
             2\Omega_{\rm LNRF} \xi^\idxII \eta_\idxII -
             \Omega^2_{\rm LNRF} \xi^\idxII \xi_\idxII
       \right).
\end{array}
\ee
The  LNRF vector field  $n^\idxI$ can  be used  for the  definition of
inertial  forces introduced  above.  Assuming a  circular motion  with
angular velocity $\Omega = \d\phi/\d  t$ as measured by the stationary
observers at infinity, the 4-velocity is given by
\be                                                
  u^\idxI = A (\eta^\idxI + \Omega \xi^\idxI),                   \quad
  A = \left(-\eta^\idxII \eta_\idxII -
              2 \Omega \xi^\idxII \eta_\idxII -
              \Omega^2 \xi^\idxII \xi_\idxII
  \right)^{-1/2};
\ee
now  $\tau^\idxI$  is directed  along  the  rotational Killing  vector
$\xi^\idxI$.   The  gravitational  (\ref{gf}),   Coriolis\nd  Lense\nd
Thirring  (\ref{cltf}),  and  centrifugal  (\ref{cf})  forces  can  be
written down as
\bea                                               
  &&G_\idxI =
    -m \Phi\!_{,\idxI} =
    -m\frac{1}{2}\partial_\idxI
    \left[\ln\left(\frac{g^2_{t\phi} -
      g_{tt}g_{\phi\phi}}{g_{\phi\phi}}\right)
    \right], \\
  &&C_\idxI =
    mA^2 \frac{\Omega - \Omega_{\rm LNRF}}{\xi^\idxII \xi_\idxII}
    \left[
      (\xi^\idxII \xi_\idxII)(\eta^\idxIII \xi_\idxIII)_{,\idxI} -
      (\eta^\idxIII \xi_\idxIII)(\xi^\idxII \xi_\idxII)_{,\idxI}
    \right] =  \nonumber \\
  &&\hphantom{C_\idxI = {}}%
    mA^2 (\Omega - \Omega_{\rm LNRF}) \sqrt{g_{\phi\phi}}
    \left[
      \partial_\idxI\left(g_{t\phi}g_{\phi\phi}^{-1/2}\right) +
      \Omega_{\rm LNRF}\,\partial_\idxI \sqrt{g_{\phi\phi}}
    \right],                                            \label{e22} \\
  &&Z_\idxI =
    \frac{1}{2} mA^2
    \frac{(\Omega-\Omega_{\rm LNRF})^2}{\iota^\idxII \iota_\idxII}
    \left[
      (\iota^\idxII \iota_\idxII)(\xi^\idxIII \xi_\idxIII)_{,\idxI} -
      (\xi^\idxIII \xi_\idxIII)(\iota^\idxII \iota_\idxII)_{,\idxI}
    \right] =  \nonumber \\
  &&\hphantom{Z_\idxI = {}}-\frac{1}{2} mA^2
    (\Omega - \Omega_{\rm LNRF})^2
    g_{\phi\phi}\,
    \partial_\idxI
    \left[
      \ln\left(
           \frac{g^2_{\phi\phi}}{g^2_{t\phi}-g_{tt}g_{\phi\phi}}
         \right)
    \right],                                               \label{e23}
\eea
respectively;  we denote  $\iota^\idxI  = \expon{-\Phi}n^\idxI$.   The
Euler  force will  appear  for $\Omega\neq  {\rm  const}$ only,  being
determined  by $\dot{\Omega} =  u^\idxII \nabla\!_\idxII  \Omega$ (see
\cite{ANW95}). The electromagnetic forces acting in charged spacetimes
were   defined  in   the  framework   of  the   optical   geometry  in
\cite{ACNSV,SA}. However,  here we  shall concentrate on  the inertial
forces  acting  on the  motion  in  the  equatorial plane  ($\theta  =
\pi/2$), and their relation to the embedding diagram of the equatorial
plane of the optical geometry.

Clearly, by definition, the  gravitational force is independent of the
orbiting particle's velocity. On  the other hand, both the Coriolis\nd
Lense\nd Thirring  and the  centrifugal force vanish  (at any  $r$) if
$\Omega  =  \Omega_{\rm LNRF}$,  i.e.,  if  the  orbiting particle  is
stationary at the  LNRF located at the radius  of circular orbit; this
fact clearly  illustrates that  the LNRF are  properly chosen  for the
definition of the optical geometry  and inertial forces in accord with
Newtonian intuition.   Moreover, both the forces vanish  at some radii
independently of  $\Omega$. In the  case of centrifugal  force, namely
this property  will be imprinted  into the structure of  the embedding
diagrams.

\section{Stationary equatorial circular motion in the Kerr\nd Newman
  spacetimes} \label{secm}

For the stationary circular motion, $\dot{\Omega}=0$. It is convenient to
express the inertial forces in terms of `Newtonian' velocity
\be
  \tilde{v} = \gamma v.
\ee
Velocity $\tilde{v}$ takes values from $-\infty$ to
$\infty$, while $v$ from $-1$ to 1. In the stationary and axially symmetric
spacetimes, there is
\be
  v = \tilde{\Omega}\tilde{R},
\ee
where
\be
  \tilde{\Omega} = \Omega - \Omega_{\rm LNRF},\quad
  \tilde{R} = \tilde{r}\expon{\Phi},\quad
  \tilde{r} =(\xi^\idxI \xi_\idxI)^{1/2}.                   \label{tilOmRr}
\ee
$\tilde{R}$ is so called radius of gyration, because $\tilde{R}^2 =
\tilde{\ell}/\tilde{\Omega}$, where $\tilde{\ell} = L/E$ is the specific
angular momentum, $L = U^\idxI \xi_\idxI$ is the angular momentum, and $E =
-U^\idxI \eta_\idxI = \gamma\expon{\Phi}$ is the
energy~\cite{ANW95,AMS}. It plays a very important role in theory of
rotational effects in strong gravitational fields. The direction of
increase of the radius of gyration gives a preferred determination of the
local outward direction relevant for the dynamical effects of rotation.
This direction becomes misaligned with the `global' outward direction in
strong fields~\cite{AMS}. The condition $\tilde{R} = {\rm const}$ defines
the von Zeipel cylinders in stationary spacetimes, which are related to the
equipotential surfaces of equilibrium configurations of perfect
fluid~\cite{KJA}.

It is convenient to introduce the gravitational acceleration, and the
velocity independent parts of the Coriolis and centrifugal accelerations in
the direction $e_\idxI$ by the relations~\cite{ANW95}
\be
  {\cal G}(r) = e^\idxI \nabla\!_\idxI \Phi,\quad
  {\cal C}(r) = e^\idxI \tilde{R}\,\nabla\!_\idxI \Omega_{\rm LNRF},\quad
  {\cal Z}(r) = e^\idxI \tilde{R}^{-1} \nabla\!_\idxI \tilde{R}.   \label{gcz}
\ee
The acceleration necessary to keep a particle in a stationary motion with a
velocity $\tilde{v}$ along a circle $r = {\rm const}$ in the equatorial
plane can then be expressed in a very simple form
\be
  a(\tilde{v},r) =
    -{\cal G}(r) - \tilde{v}^2 {\cal Z}(r) +
    (1+\tilde{v}^2)^{1/2}\tilde{v}{\cal C}(r)                  \label{accel}
\ee
that enables an effective discussion of the properties of both accelerated
and geodesic motion. (Of course, only the positive root of the last term on
the r.h.s.\ of Eq.~(\ref{accel}) has physical meaning.)

For the stationary circular motion in the equatorial plane of the Kerr\nd
Newman spacetimes all three parts of the acceleration have only radial
components. We obtain
\bea
  {\cal G}(r) &=&
    -\frac{r^4(r-e^2)+a^2[2r(r-2)(r-e^2)-e^4]+a^4(r-e^2)}%
          {r\Delta [(r^2+a^2)^2 - a^2\Delta]}, \\
  {\cal C}(r) &=&
    -\frac{2a[r(3r^2+a^2)-e^2(2r+a^2)]}%
         {r\sqrt{\Delta}\,[(r^2+a^2)^2-a^2\Delta]}, \\
  {\cal Z}(r) &=&
    \left\{r\Delta [(r^2+a^2)^2-a^2\Delta]\right\}^{-1}
    \left\{r^4(r^2-3r+2e^2)+\right.                     \nonumber \\
    &&\left.a^2[r^2(r^2-3r+6)+e^2r(3r-7)+2e^4]-2a^4(r-e^2)\right\}. \label{zr}
\eea
These definitions of the gravitational and inertial forces have really a
Newtonian character, since the gravitational force $G=-m{\cal G}(r)$ is
velocity independent, $C = m(1+v^2)^{1/2}\tilde{v}{\cal C}(r)$ depends on
$\tilde{v}$, and the centrifugal force $Z = -m\tilde{v}^2 {\cal Z}(r)$
depends on $\tilde{v}^2$. Moreover, asymptotic behaviour of these forces is
consistent with `Newtonian' intuition:
\be
  {\cal G}(r\rightarrow\infty)\sim -\frac{1}{r^2},\quad
  {\cal C}(r\rightarrow\infty)\sim -\frac{a}{r^3},\quad
  {\cal Z}(r\rightarrow\infty)\sim  \frac{1}{r}.               \label{asymbeh}
\ee 

It follows immediately from Eq.~(\ref{accel}) that the photon circular
geodesic motion ($\tilde{v}^2\rightarrow\infty$) is determined by the
conditions
\bea
  &&{\cal Z}(r) - {\cal C}(r) = 0, \qquad \mbox{corotating orbits}, \\
  &&{\cal Z}(r) + {\cal C}(r) = 0, \qquad \mbox{counterrotating orbits}.
\eea
For ultrarelativistic particles ($\tilde{v}\gg 1$, $\tilde{v}\ll -1$), we
  obtain an asymptotic relation
\be
  a(\tilde{v},r) \approx
    -{\cal G}(r) \pm {\textstyle\frac{1}{2}}{\cal C}(r) -
    v^2 [{\cal Z}(r)\mp {\cal C}(r)].
\ee
The upper signs correspond to the corotating motion ($\tilde{v}>0$), the
lower signs to the counterrotating ($\tilde{v}<0$) motion. The
ultrarelativistic particles moving on the radius of the corotating
(counterrotating) photon circular geodesic are kept by the acceleration
\be
  a(r) = -{\cal G}(r) \pm {\textstyle\frac{1}{2}}{\cal C}(r),
\ee
which is independent of velocity, with accuracy $O(\tilde{v}^{-2})$. In
static spacetimes ${\cal C}(r) = 0$, and at the radius of the photon circular
orbit the acceleration of particles is
\be
  a(r) = -{\cal G}(r),                                          \label{aEQ-g}
\ee
and is independent of $\tilde{v}$ exactly. (The relation (\ref{aEQ-g}) is
not limited to the case of ultrarelativistic orbits, because ${\cal Z}(r) =
0$ at the radius of the photon circular geodesics in static spacetimes.)

Velocities of particles moving along the circular geodesics are determined
by the relation
\be
  \tilde{v}^2_\pm =
    \frac{\frac{1}{2}{\cal C}^2 - {\cal Z}{\cal G} \mp \frac{1}{2}{\cal
        C}({\cal C}^2-4{\cal Z}{\cal G}+4{\cal G}^2)^{1/2}}%
         {{\cal Z}^2-{\cal C}^2},
\ee
where the upper (lower) signs correspond to the corotating (counterrotating)
orbits, $\tilde{v}>0$ ($\tilde{v}<0$).

\begin{figure}[b]
\begin{minipage}[b]{.48\hsize}
\centering \leavevmode \epsfxsize=\hsize \epsfbox{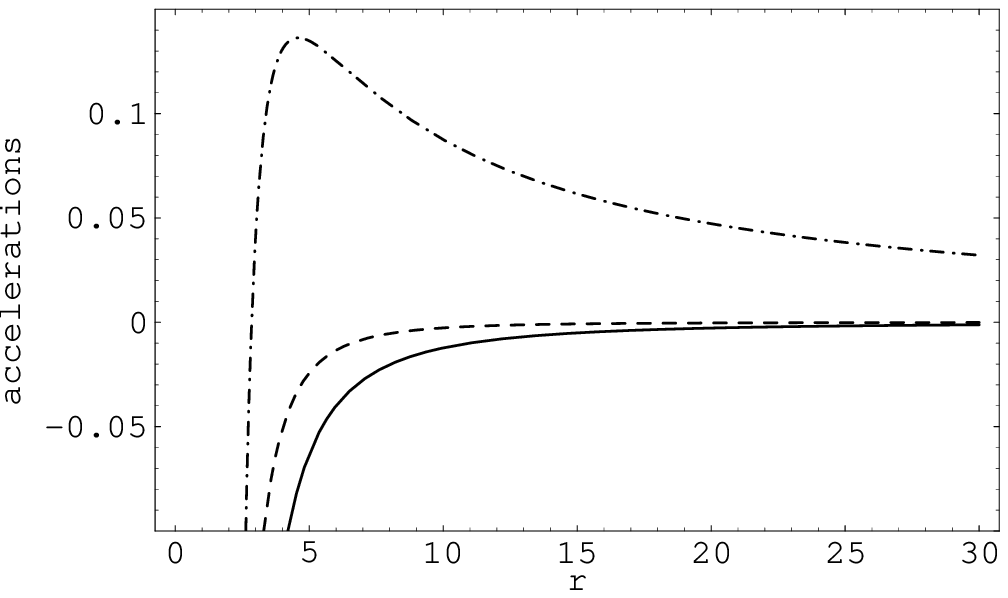}
\par\vskip 1mm {\small (a)}
\end{minipage}\hfill%
\begin{minipage}[b]{.48\hsize}
\centering \leavevmode \epsfxsize=\hsize \epsfbox{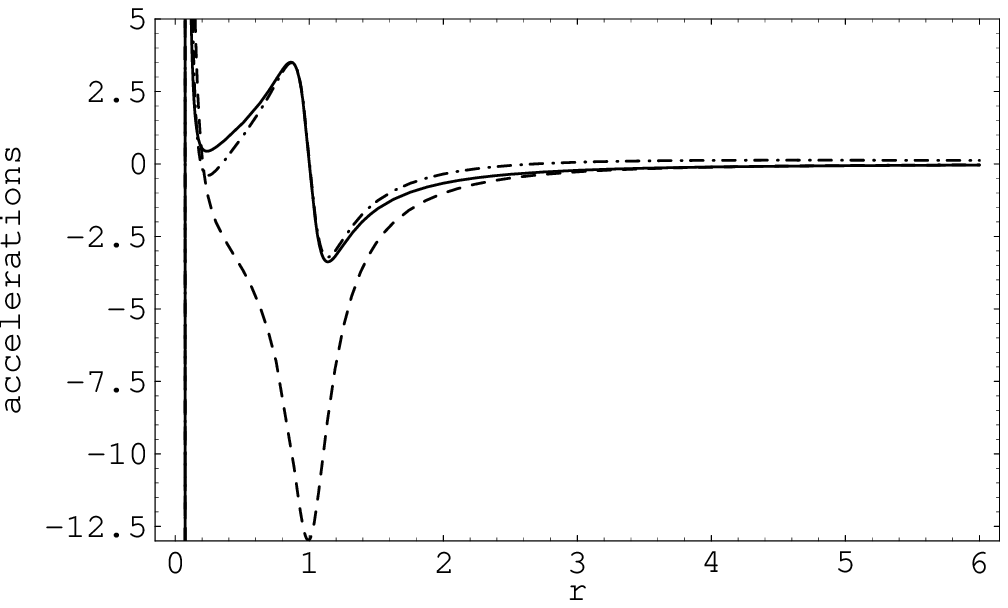}
\par\vskip 1mm {\small (b)}
\end{minipage}
\caption{Behaviour of the velocity-independent parts of gravitational
  (solid), Coriolis (dashed), and centrifugal (dashed-dotted) forces ${\cal
    G}(r)$, ${\cal C}(r)$, and ${\cal Z}(r)$, respectively. (a)~$a^2=0.16$,
  $e^2=0.16$, (b)~$a^2=0.86$, $e^2=0.16$.}
\label{addf1}
\end{figure}

Properties of the gravitational acceleration ${\cal G}(r)$, and of the
velocity independent parts of the Coriolis and centrifugal of acceleration
${\cal C}(r)$ and ${\cal Z}(r)$ determine properties of both the
accelerated, and geodesic motion in a given spacetime. It is clear that for
the geodesic circular motion at a given $r$, the condition
\be
  {\cal C}^2(r) - 4{\cal Z}(r){\cal G}(r) + 4{\cal G}^2(r) > 0
\ee
must be satisfied. In static spacetimes ${\cal C}(r) = 0$, and the reality
condition of the geodesic orbits is ${\cal G}(r){\cal Z}(r)<0$, i.e., the
gravitational and centrifugal forces must point in opposite directions. If
${\cal C}(r)\neq 0$, the cases $|{\cal Z}|>|{\cal C}|$, $|{\cal Z}|=|{\cal
  C}|$, $|{\cal Z}|<|{\cal C}|$ give qualitatively different situations. Of
course, the gravitational force can play an important role too. Therefore,
we have to determine behaviour of the functions ${\cal G}(r;a,e)$, ${\cal
  C}(r;a,e)$, and ${\cal Z}(r;a,e)$. We give an illustration of the
behaviour of these functions for a black-hole (Fig.\,\ref{addf1}a) and
naked-singularity (Fig.\,\ref{addf1}b) spacetimes. The asymptotic behaviour
of these functions is given by the relations (\ref{asymbeh}). Further, it
is important to know, if these functions are positive or negative.

We have to find, where the functions ${\cal G}(r;a,e)$, ${\cal C}(r;a,e)$,
and ${\cal Z}(r;a,e)$ change their sign. All of these functions diverge,
and change their sign, at the boundary of the region of causality
violations, which is determined by the relations
\be
  a^2 = a^2_{\rm c.v.}(r;e) = \frac{r^4}{e^2 - 2r - r^2}.
\ee
In the Kerr spacetimes ($e^2=0$), this region is restricted to $r<0$.

Surprisingly, the gravitational acceleration changes its sign at the zero
points given by
\be
  a^2 = a^2_{{\rm g}\pm}(r;e) =
    \frac{e^4 - 2r(r-2)(r-e^2) \pm \sqrt{D_{\rm g}}}{2(r-e^2)},
\ee
with
\be
  D_{\rm g} = 4r(r-2)(r-e^2)(2re^2 - 2r^2 - e^4) + e^8.
\ee
It holds at $r>0$ even for the Kerr spacetimes---we arrive at a
simple formula
\be
  a^2 = a^2_{{\rm g}\pm}(r) = r (2 - r \pm \sqrt{2-r}).
\ee

The Coriolis acceleration changes sign at zero points determined by
\be
  a^2 = a^2_{\rm c}(r;e) = \frac{r(3r^2 - 2e^2)}{e^2 - r}.
\ee
In the Kerr spacetimes, ${\cal C}(r)$ does not change sign at $r>0$.

\begin{figure}[b]
\centering \leavevmode \epsfxsize=.7\hsize \epsfbox{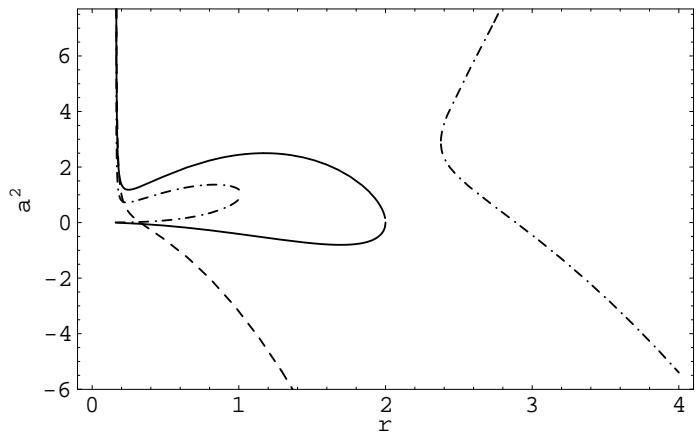}
\caption{Behaviour of the functions $a_{{\rm g}\pm}^2$ (solid),
  $a_{\rm c}^2$ (dashed), and $a_{{\rm z}\pm}^2$ (dashed-dotted). They are
  given for fixed charge parameter of the spacetime $e^2 = 0.16$.}
\label{addf2}
\end{figure}

Finally, we find that the centrifugal acceleration changes sign at radii
given by
\be                                                  
  a^2 = a^2_{{\rm z}\pm}(r; e) =
    \frac{r^2 (r^2-3r+6) +e^2 r (3r-7) + 2 e^4 \pm \sqrt{D_{\rm z}}}%
    {4(r-e^2)},                                            \label{e26}
\ee
where the discriminant is
\bea                                                 
  D_{\rm z}(r;e) &=&
    r^8+2r^7 - (3+2 e^2) r^6 + 4 (2e^2-9)r^5 +\nonumber\\
    &&3(12+26 e^2-e^4)r^4 - \nonumber \\
    &&2 e^2(42+27e^2)r^3 +
      e^4(73+12e^2)r^2 - 28e^6 r + 4 e^8.                  \label{e27}
\eea
We give an example of the behaviour of all the functions $a_{{\rm
    g}\pm}^2(r;e)$, $a_{{\rm c}}^2(r;e)$, and $a_{{\rm z}\pm}^2(r;e)$ in
Fig.\,\ref{addf2}. Generally, properties of the stationary circular motion
can be given in a straightforward manner by using these functions; we shall
not discuss details here.

The centrifugal acceleration is very important, since it is closely related
to the radius of gyration (see Eq.~(\ref{gcz})). Therefore, it deserves a
detailed study.

\begin{figure}[t]
\centering \leavevmode \epsfxsize=.7\hsize \epsfbox{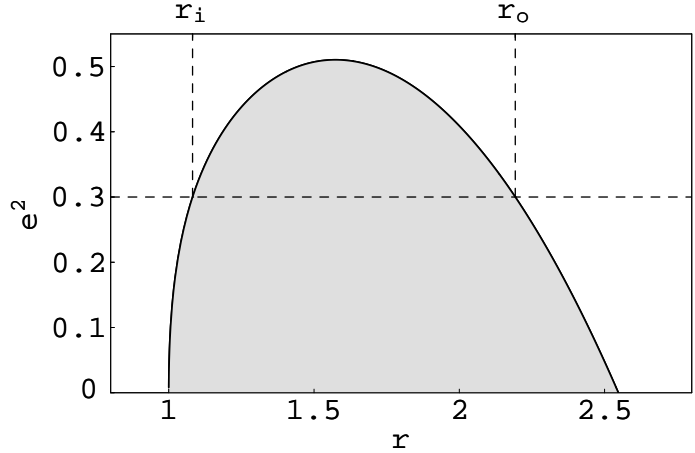}
\caption{The function determining reality condition $D_{\rm z}(r;e)\geq 0$ for
  the radii at which the centrifugal force vanishes. Inside the shaded
  region, there is $D_{\rm z}(r;e)<0$, and the functions $a^2_{{\rm
      z}\pm}(r;e)$ are not defined there. The three remaining solutions of
  the equation $D_{\rm z}(r;e)=0$ are not presented here, being physically
  irrelevant (with $a^2<0$ or $e^2<0$).}
\label{r-e2}
\end{figure}

The reality condition for $a^2_{{\rm z}\pm}(r;e)$ is
\be                                                    
  D_{\rm z}(r;e) \geq 0,                                           \label{e28}
\ee
which can be treated easily  considering $D_{\rm z}(r;e)$ as a quartic polynom
in $e^2$. The results are given in Fig.\,\ref{r-e2}.  Detailed behaviour of
the functions  $a^2_{{\rm z}\pm}(r;e)$ will be determined  in the next
section, since it is closely related to embedding diagrams of the
equatorial plane of the optical geometry.

\section{Embedding diagrams of the optical geometry}      \label{edog}

Metric  coefficients of  the optical  geometry of  the  Kerr\nd Newman
spacetimes are, due to Eq.~(\ref{e9}), given by the relations
\be                                               
  \tilde{h}_{rr} =
    \frac{\tilde{h}_{\theta\theta}}{\Delta} =
    \frac{A}{\Delta^2},  \quad
  \tilde{h}_{\phi\phi} =
    \frac{A^2}{\Delta\Sigma^2}\sin^2\theta.
\ee
In the equatorial plane they reduce to
\[
  \tilde{h}_{rr}(\theta={\textstyle\frac{1}{2}}\pi) =
    \frac{(r^2+a^2)^2 -a^2 \Delta}{\Delta^2},
\]
\be
  \tilde{h}_{\phi\phi}(\theta = {\textstyle\frac{1}{2}}\pi) =
    \frac{\left[(r^2+a^2)^2 -a^2
    \Delta \right]^2}{r^4 \Delta}.                         \label{e30}
\ee

The properties of  the optical geometry related to  the Kerr\nd Newman
spacetimes  can  conveniently  be  represented  by  embedding  of  the
equatorial  (symmetry) plane  into the  3-dimensional  Euclidean space
with   line   element  expressed   in   the  cylindrical   coordinates
($\rho,z,\phi$) in the form
\be                                                 
  \d\sigma^2 = \d\rho^2 + \rho^2\,\d\phi +\d z^2.
\ee
The  embedding  diagram  is  characterised by  the  embedding  formula
$z=z(\rho)$ determining a surface in the Euclidean space with the line
element
\be                                              
  \d\ell^2_{\rm (E)} =
    \left[1+\left(\oder{z}{\rho}\right)^2 \right]\d\rho^2 +
    \rho^2\,\d\phi^2
\ee
isometric to  the 2-dimensional equatorial plane of  the optical space
determined by the line element
\be                                               
  \d\tilde{\ell}^2 =
    \tilde{h}_{rr}\,\d r^2 +
    \tilde{h}_{\phi\phi}\,\d\phi^2
\ee
with $\tilde{h}_{rr}$ and $\tilde{h}_{\phi\phi}$ given by (\ref{e30}).

The azimuthal  coordinates can  be identified, the  radial coordinates
are related as
\be      
  \rho^2 = \tilde{h}_{\phi\phi},                           \label{e35}
\ee
and the embedding formula is governed by the relation
\be                                         
  \left(\oder{z}{\rho} \right)^2 =
    \tilde{h}_{rr} \left(\oder{r}{\rho}\right)^2 - 1.
\ee
It is convenient  to transfer the embedding formula  into a parametric
form $z(\rho) = z(r(\rho))$ with $r$ being the parameter. Then
\be                           
  \oder{z}{r} =
    \pm\sqrt{\frac{(r^2+a^2)^2 - a^2\Delta}{\Delta^2} -
    \left(\oder{\rho}{r}\right)^2}.                        \label{e37}
\ee
The sign in this formula is irrelevant, leading to isometric surfaces.
Because
\be                                         
  \oder{z}{\rho} = \oder{z}{r}\oder{r}{\rho},
\ee
the turning  points of the  embedding diagram, giving its  throats and
bellies, are determined by the condition $\d\rho/\d r = 0$, where
\bea                                             
  \oder{\rho}{r} &=& \{r^4 [r(r-3)+2 e^2] + a^2
      [2e^4+ e^2 r(3r-7)+ r^2(r^2-3r+6)]-                 \nonumber \\
    &&2a^4(r- e^2)\}
    [r^3 (r^2-2r+a^2+ e^2)^{3/2}].                      \label{drhodr}
\eea
By comparing Eqs~(\ref{zr}) and (\ref{drhodr}) we can immediately see
that turning  points of the  embedding diagrams are really  located at
the radii where the centrifugal force vanishes and changes sign. Thus,
we can conclude  that just this property of  embeddings of the optical
geometry   of  the  vacuum   spherically  symmetric   spacetimes  (see
\cite{ACL,KSA,SH99b})     survives    in     the     Kerr\nd    Newman
spacetimes.  However, photon  circular orbits  are displaced  from the
radii   corresponding  to   the  turning   points  of   the  embedding
diagrams. Therefore,  it is interesting  to find the  situations where
the photon circular orbits lie  within the regions of embeddability of
the optical geometry.

\subsection{Limits of embeddability}

The embeddability  condition $\left(\d z/\d  r \right)^2 \geq  0$ (see
(\ref{e37})) implies the relation
\bea
  E(r;a,e) &=&
    4r^{11}-3(e^2+3)r^{10}+12(a^2+e^2)r^9-                 \nonumber\\
  &&2(17a^2+5a^2e^2+2e^4)r^8+4(9a^2+3a^4+12a^2 e^2)r^7-    \nonumber\\
  &&(33a^4+66a^2 e^2+ 11a^4 e^2+ 17a^2 e^4)r^6+            \nonumber\\
  &&4(9a^4+a^6+13a^4e^2+10a^2e^4)r^5-                      \nonumber\\
  &&(36a^4+12a^6+78a^4e^2+4a^6e^2+21a^4e^4+8a^2e^6)r^4+    \nonumber\\
  &&2(12a^6+42a^4e^2+12a^6e^2+27a^4 e^4)r^3-               \nonumber\\
  &&(4a^8+ 52a^6 e^2+73a^4e^4+12a^6e^4+12a^4e^6)r^2+       \nonumber\\
  &&4(2a^8e^2+9a^6e^4+7a^4e^6)r-                           \nonumber\\
  &&4(a^8e^4+2a^6e^6+a^4e^8)
    \geq 0.
\eea
The function $E(r;a,e)$ can be considered as a polynom quartic both in
$a^2$  and $e^2$.  For $e^2  = 0$,  the function  $E(r;a)$ is  still a
polynom quartic in $a^2$  (see \cite{SH99c} for details). However, for
$a^2 = 0$, the  function $E(r;e)$ simplifies significantly, being only
a polynom quadratic in $e^2$
\be                                                   
  E(r;e) = r^8 [r^2(4r-9) + 3r(4-r)e^2 - 4e^4];            \label{e41}
\ee
its behaviour is discussed in  \cite{KSA}. For $a^2=e^2=0$ we arrive at
the well-known \Sch\ condition $r \geq\frac{9}{4}$.

The  limits   of  embeddability   are  determined  by   the  condition
$E(r;a,e)=0$ which can  be solved as a quartic  equation in $a^2$. Let
us   denote   the  four   solutions   as   $a^2_{\e  k}(r;e)$,   $k\in
\{1,2,3,4\}$. Instead of giving long explicit expressions for the four
solutions  $a^2_{\e  k}(r;e)$  we  will  treat  them  numerically  and
classify  qualitatively; different  types of  their behaviour  will be
described.   Of  course,  we   naturally  restrict  our  attention  to
physically relevant situations when $a^2 \geq 0$ and $e^2 \geq 0$.

\begin{figure}[b]
\centering \leavevmode \epsfxsize=.6\hsize \epsfbox{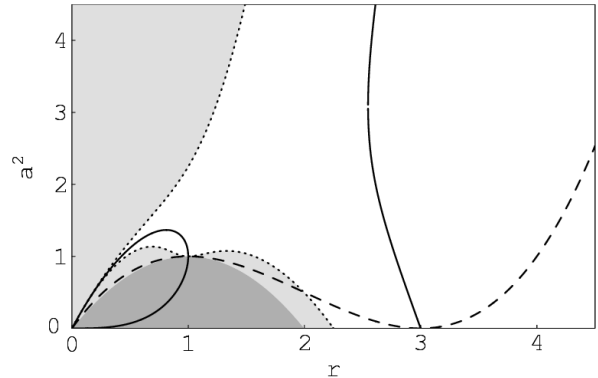}
\caption{Square of   the  specific   angular  momentum  of   the  Kerr
  backgrounds corresponding to the event horizons ($a^2_{\rm h}(r;e)$,
  the border between the dark  gray dynamic area, in which the optical
  geometry is  not defined, and  the light gray  non-embeddable area),
  photon circular orbits  ($a^2_{\rm ph}(r;e)$, dashed curve), turning
  points  of  the embedding  diagram  ($a^2_{{\rm z}\pm}(r;e)$,  solid
  curve), and embeddability  region border ($a^2_{\rm e}(r;e)$, dotted
  curve enclosing  the light gray  area not embeddable  into Euclidean
  space)  are  drawn  as  functions  of the  radius  for  fixed  value
  $e^2=0$.}
\label{r-a2-000}
\end{figure}

Recall that if $e^2 = 0$, the limits of embeddability are given by the
solutions  $a^2_{\e3}(r)$, $a^2_{\e4}(r)$  representing  two branches,
both  reaching zero  at  $r=0$ (see  Fig.\,\ref{r-a2-000}). The  upper
branch ($a^2_{\e4}$)  diverges for $r \rightarrow  \infty$.  The lower
one ($a^2_{\e3}$)  has a local  minimum at $r=1$,  where $a^2_{\e({\rm
min})} =1$, and two local maxima, $a^2_{\e({\rm max},2)} = 1.13540$ at
$r=0.68947$, and $a^2_{\e({\rm max},1)} = 1.07543$ at $r=1.33172$; its
second zero point  is located at $r=2.25$, corresponding  to the \Sch\
case \cite{SH99c}.

If  $e^2  > 0$,  the  limits of  embeddability  still  consist of  two
branches. We can consider three qualitatively different situations.

\subsubsection{Class Ea: {\protect\boldmath$e^2\in (0,0.68950)$}}

The  two branches  are  given by  the  solutions $a^2_{\e3}(r;e)$  and
$a^2_{\e4}(r;e)$,  again. Both  the branches  diverge at  $r=e^2$. The
upper  branch ($a^2_{\e4}$)  diverges  also for  $r\rightarrow\infty$,
having a local minimum  $a^2_{\e4({\rm min})}$ near $r=e^2$. The lower
branch  ($a^2_{\e3}$)  has always  a  local  minimum  at $r=1$,  where
$a^2_{\e({\rm min})}  = 1-  e^2$, corresponding to  extreme black-hole
states,  and a  local  maximum  $a^2_{\e({\rm max},1)}  >  1- e^2$  at
$r>1$. Its  zero point,  corresponding to the  Reissner\nd Nordstr\"om
case, is  determined by  $E(r;e) = 0$  (cf.\ Eq.~(\ref{e41})).  It can
also have a local minimum  $a^2_{\e3 ({\rm min})}$ and a local maximum
$a^2_{\e3  ({\rm  max},2)}$  at  $r<1$.  Positions  of  $a^2_{\e4({\rm
min})}$ and $a^2_{\e3({\rm min})}$  relative to the value $a^2=1-e^2$,
and the  relation between  $a^2_{\e4({\rm min})}$ and  $a^2_{\e3 ({\rm
max},2)}$, or  $a^2_{\e3 ({\rm max},1)}$,  determine different members
of   embeddable  regions   of  the   Kerr\nd  Newman   black-hole  and
naked-singularity    spacetimes.   We    can   give    the   following
subclassification of the spacetimes listed by the number of embeddable
regions  according to  the values  of the  parameter $e^2$;  for fixed
$e^2$ from given  interval, the numbers of the  embeddable regions are
given with the parameter $a^2$ growing:

\begin{figure}[p]
\centering \leavevmode \epsfxsize=.58\hsize \epsfbox{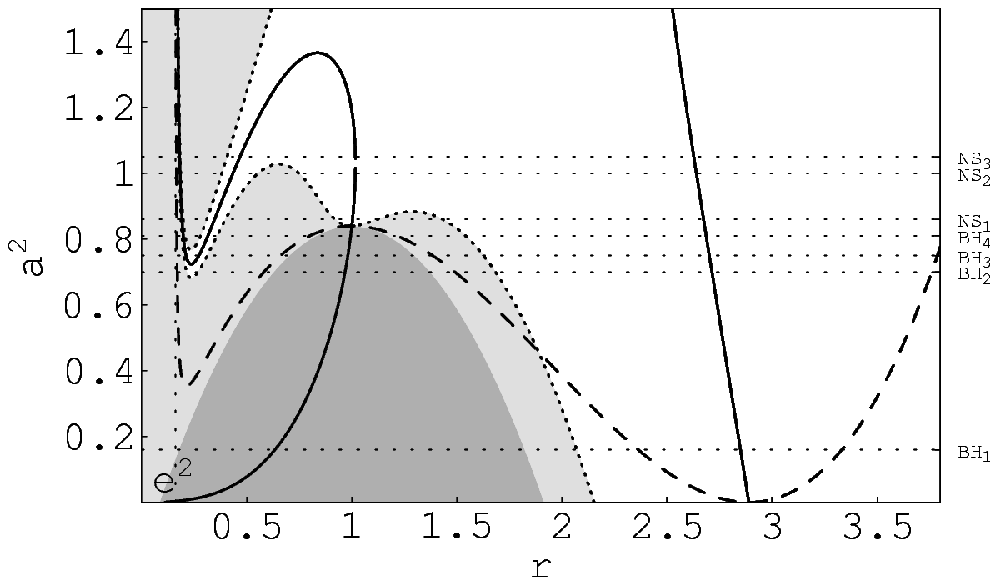}
\caption{Square of the specific angular momentum of the Kerr\nd Newman
  backgrounds corresponding to the event horizons ($a^2_{\rm h}(r;e)$,
  between the dark gray dynamic area, in which the optical geometry is
  not  defined,  and  the  light  gray  non-embeddable  area),  photon
  circular orbits ($a^2_{\rm  ph}(r;e)$, dashed curve), turning points
  of the embedding diagram ($a^2_{{\rm z}\pm}(r;e)$, solid curve), and
  embeddability  region  border   ($a^2_{\rm  e}(r;e)$,  dotted  curve
  enclosing the  light gray area not embeddable  into Euclidean space)
  are   drawn   as  functions   of   the   radius   for  fixed   value
  $e^2=0.16$. This case covers  all four types of black-hole embedding
  diagrams   and   the   first   three  cases   of   naked-singularity
  diagrams. They  are determined by  the lines $a^2={\rm  const}$, and
  depicted  by the  notation of  the classification  according  to the
  properties  of  embedding   diagrams  (BH$_1$\nd  BH$_4$,  NS$_1$\nd
  NS$_3$).  The  dotted vertical line is common  vertical asymptote of
  $a^2_{\rm   ph}(r;e)$,   $a^2_{{\rm   c}\pm}(r;e)$,  and   $a^2_{\rm
  e}(r;e)$.}
\label{r-a2-016}
\end{figure}

\begin{figure}[p]
\centering \leavevmode \epsfxsize=.58\hsize \epsfbox{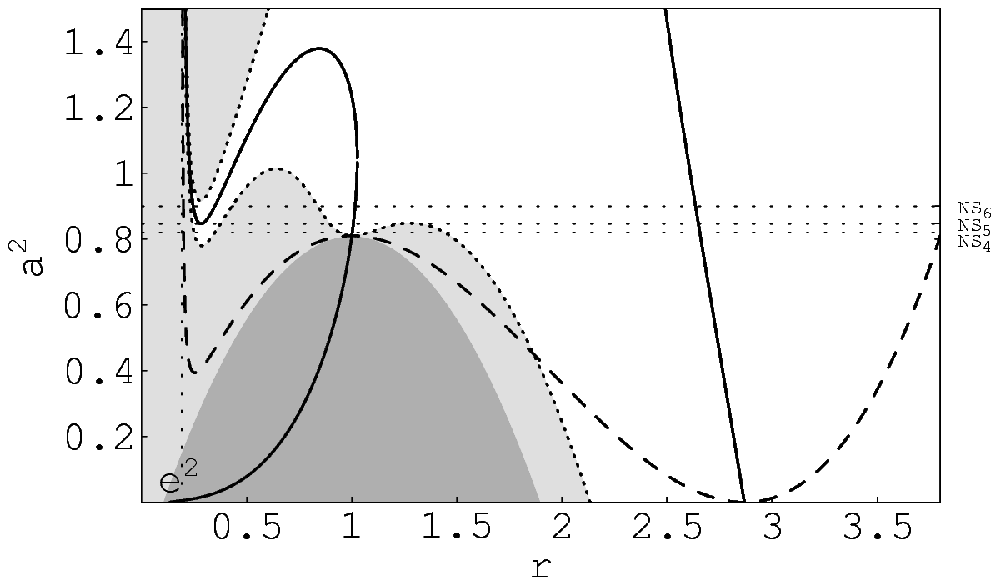}
\caption{The  functions  $a^2_{\rm   h}(r;e)$,   $a^2_{\rm  ph}(r;e)$,
  $a^2_{{\rm  c}\pm}(r;e)$,  and $a^2_{\rm  e}(r;e)$  drawn for  fixed
  value  $e^2=0.19$.  The  naked-singularity classes  NS$_4$\nd NS$_6$
  are depicted.}
\label{r-a2-019}
\end{figure}

\begin{figure}[p]
\centering \leavevmode \epsfxsize=.58\hsize \epsfbox{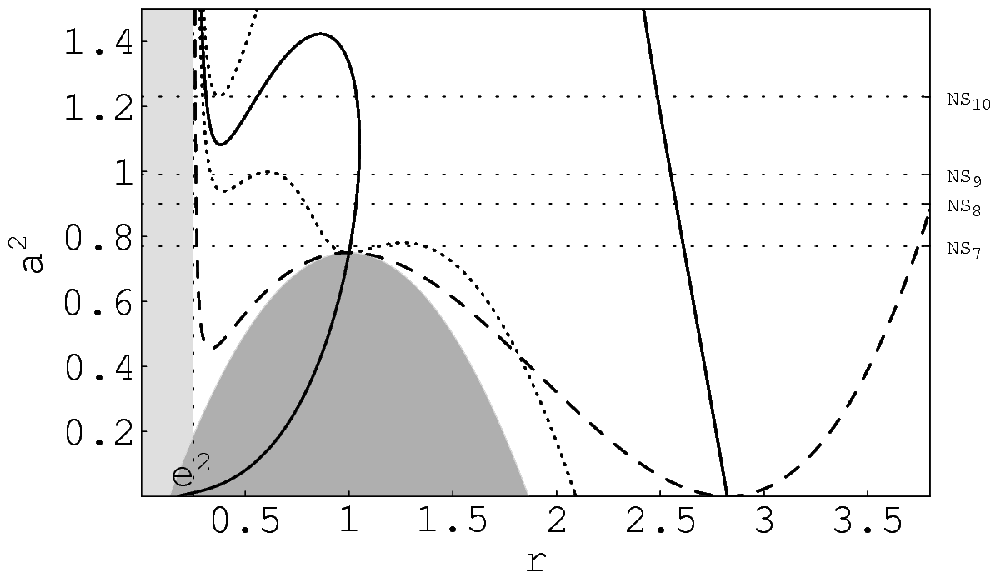}
\caption{The  functions  $a^2_{\rm   h}(r;e)$,   $a^2_{\rm  ph}(r;e)$,
  $a^2_{{\rm  c}\pm}(r;e)$,  and $a^2_{\rm  e}(r;e)$  drawn for  fixed
  value $e^2=0.25$.  The naked-singularity classes NS$_7$\nd NS$_{10}$
  are depicted.}
\label{r-a2-025}
\end{figure}

\begin{center}
\begin{tabular}{llll}
\hline
Subclass & Interval of $e^2$      & Black holes & Naked singularities\\ \hline 
Ea$_1$   & $\bra0,      0.17230)$ & $1,2,3$     & $4,3,2$            \\
Ea$_2$   & $\bra0.17230,0.17906)$ & $1,2$       & $3,4,3,2$          \\
Ea$_3$   & $\bra0.17906,0.19749)$ & $1,2$       & $3,2,3,2$          \\
Ea$_4$   & $\bra0.19749,0.20670)$ & 1           & $2,3,2,3,2$        \\
Ea$_5$   & $\bra0.20670,0.20794)$ & 1           & $2,1,2,3,2$        \\
Ea$_6$   & $\bra0.20794,0.28961)$ & 1           & $2,1,2,1,2$        \\
Ea$_7$   & $\bra0.28961,0.68950)$ & 1           & $2,1,2$            \\ \hline
\end{tabular}
\end{center}
Notice that in all of the subclasses Ea$_1$--Ea$_7$, both the branches
of the limit of embeddability are  defined at $r > e^2$. The behaviour
of  the embeddability  limits in  the subclasses  Ea$_1$,  Ea$_3$, and
Ea$_6$ is illustrated  in the Figs~\ref{r-a2-016}, \ref{r-a2-019}, and
\ref{r-a2-025}, respectively.

\subsubsection{Class Eb: {\protect\boldmath$e^2\in (0.68950,2.205)$}}

The  limits of  embeddability  have  two branches.  The  upper one  is
determined  by  the solution  $a^2_{\e4}(r;  e)$.  Again,  it is  well
defined at $r> e^2$ only, diverges at $r = e^2$ and for $r \rightarrow
\infty$, having a minimum near $r= e^2$. The lower branch is radically
different from the class  Ea.  The solution $a^2_{\e3}(r; e)$ diverges
at $r=e^2$ again,  but it has a discontinuity,  which is `filled up'
by   various  combinations   of  the   solutions   $a^2_{\e1}(r;  e)$,
$a^2_{\e2}(r; e)$,  and $a^2_{\e4}(r; e)$. The  combinations depend on
the  parameter  $e^2$.  However,  it  is not  worth  to  discuss  them
explicitly  because they have  common basic  properties. They  have no
local  extrema,  but  two  lobes---an  external (upper)  one,  and  an
internal (lower) one.  The internal lobe can enter the  region of $r <
e^2$ for $e^2$ high enough; for $e^2 >1$, the internal lobe is shifted
to the physically irrelevant region,  where $a^2 <0$. If $e^2 <1$, the
solution $a^2_{\e3}(r; e)$  has a local minimum at  $r=1$, with $a^2 =
1- e^2$  corresponding to an  extreme black-hole, and a  local maximum
$a^2_{\e({\rm max},1)}  > 1-e^2$ at $r>1$;  for $e^2 =  1$ these local
extrema   coalesce   at    $r=1$,   $a^2_{\e3}=0$.    Therefore,   the
subclassification according to the number of embeddable regions can be
given in the following simple way (with $a^2$ growing):

\begin{figure}[t]
\centering \leavevmode \epsfxsize=.6\hsize \epsfbox{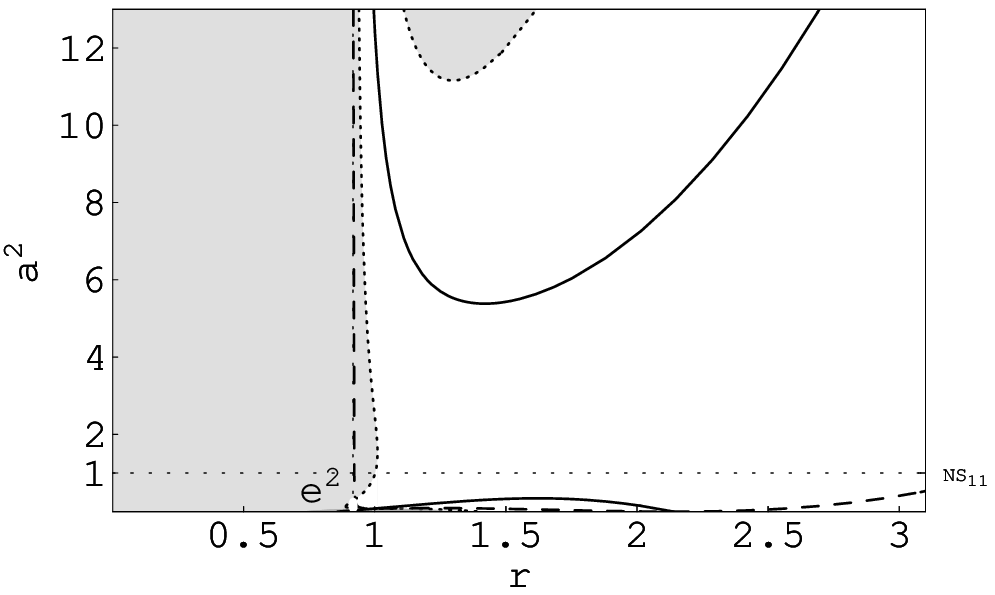}
\par\vskip 2mm\par
\centering \leavevmode \epsfxsize=.6\hsize \epsfbox{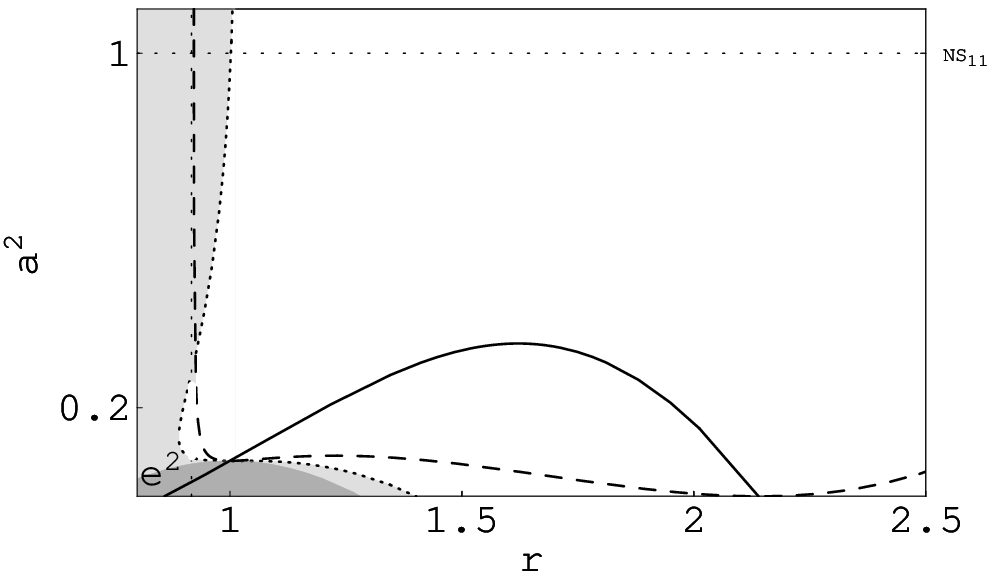}
\caption{The  functions  $a^2_{\rm   h}(r;e)$,   $a^2_{\rm  ph}(r;e)$,
  $a^2_{{\rm  c}\pm}(r;e)$,  and $a^2_{\rm  e}(r;e)$  drawn for  fixed
  value  $e^2=0.92$.  Contrary  to  the preceding  cases, there  exist
  lobes of $a^2_{\rm e}(r;e)$,  and $a^2_{{\rm z}\pm}(r;e)$ is defined
  for  all  $r\geq e^2$.  However,  there  exist  an interval  of  the
  parameter  $a^2$ corresponding  to naked-singularity  spacetimes, at
  which no turning points appear on the embedding diagrams---these are
  the NS$_{11}$-spacetimes.}
\label{r-a2-090}
\end{figure}

\begin{center}
\begin{tabular}{llll}
\hline
Subclass & Interval of $e^2$ & Black holes &  Naked singularities \\ \hline 
Eb$_1$   & $\bra0.68950,1)$  & 1           &  $2,1,2$             \\
Eb$_2$   & $\bra1,2.205)$    & none        &  $1,2$               \\ \hline
\end{tabular}
\end{center}
An example of subclass Eb$_1$ can be seen in Fig.\,\ref{r-a2-090}.

\subsubsection{Class Ec: {\protect\boldmath$e^2\in (2.205,\infty)$}}

The limits of embeddability $a^2_{\e}(r; e)$ have two branches, again.
However,  now  both  the  branches  are  determined  by  the  solution
$a^2_{\e4}(r;e)$. The first branch  is defined at $r>e^2$, diverges at
$r=e^2$ and for $r \rightarrow \infty$, having a local minimum near $r
=  e^2$. The  second  branch is  relevant  from some  $r< e^2$,  where
$a^2_{\e4}  =0$,   and  it  diverges  at  $r=e^2$,   having  no  local
extreme. The  subclassification according to the  number of embeddable
regions is simply given just by one case (with $a^2$ growing):

\begin{center}
\begin{tabular}{llll}
\hline
Subclass & Interval of $e^2$   & Black holes & Naked singularities \\ \hline 
Ec$_1$   & $\bra2.205,\infty)$ & none        & $1,2$               \\ \hline
\end{tabular}
\end{center}

Note  that in  situations which  are necessary  in order  to construct
typical embedding  diagrams, the  curves $a^2_{\rm e}(r;e)$  are given
explicitly (see Figs~\ref{r-a2-016}--\ref{r-a2-090}). The behaviour of
the embeddability  limits of the  class Ec is not  illustrated because
they give no qualitatively different kind of embedding diagrams.

\subsection{Turning points of the embedding diagrams}

The embeddable regions are characterised  by the number of radii where
the embedding  diagrams have  turning points, or,  equivalently, where
the centrifugal  force determined by  (\ref{zr}) vanishes. Therefore,
the  turning points  of the  diagrams  are governed  by the  functions
$a^2_{{\rm z}\pm}(r;e)$ determined  by (\ref{e26}) and (\ref{e27}). We
shall  discuss  the  behaviour  of  these  functions  and  introduce  a
corresponding  classification   of  the  Kerr\nd   Newman  backgrounds
(relative to their parameter $e^2$) according to the number of turning
points of the embedding diagrams of their optical geometry.

It  follows   from  the   reality  condition  (\ref{e28})   (see  also
Fig.\,\ref{r-e2})  that for  $e^2<e^2_{\rm cr}=0.51033$  the functions
$a^2_{{\rm  c}\pm}(r; e)$ are  not defined  between the  radii $r_{\rm
i}(e)$,   $r_{\rm  o}(e)$,   i.e.,   inside  the   shaded  region   in
Fig.\,\ref{r-e2}.   If $e^2=0$  (see  Fig.\,\ref{r-a2-000}), there  is
$a^2_{{\rm  c}\pm}(r=0)=0$, and $a^2_{{\rm  c}-}(r=3)=0$. The  part of
$a^2_{{\rm  c}-}(r, 0)$  starting at  $r=0$ is  located in  the region
between  the horizons,  and,  therefore, is  physically irrelevant;  a
local maximum of $a^2_{{\rm z}+}(r)$  is located at $r_{\rm (max,1)} =
0.81159$  with  $a^2_{{\rm z}+{\rm  (max,1)}}  = 1.36668$;  $a^2_{{\rm
c}+}(r)$ and  $a^2_{{\rm z}-}(r)$  coincide at $r=1$  with $a^2  = 1$,
corresponding to an extreme black-hole (see \cite{SH99c}).

For $e^2 >0$,  we can separate qualitatively different  classes of the
behaviour of $a^2_{{\rm z}\pm}(r; e)$ in the following way.

\subsubsection{Class Ta: {\protect\boldmath$e^2\in (0,0.51033)$}}

According  to   the  reality  condition   (\ref{e28}),  the  functions
$a^2_{{\rm z}\pm}(r;e)$ are  defined at $r \leq r_{\rm  i}(e)$, and $r
\geq r_{\rm  o}(e)$. For $r=r_{\rm i}(e)$, $r=r_{\rm  o}(e)$, there is
$a^2_{{\rm z}+} = a^2_{{\rm  c}-}$.  At $r>r_{\rm o}(e)$, the function
$a^2_{{\rm z}-}(r; e)$ has a  zero point, while $a^2_{{\rm z}+}(r; e)$
diverges  for  $r \rightarrow  +  \infty$.   The functions  $a^2_{{\rm
c}\pm}(r; e)$ have  no extrema at $r > r_{\rm o}  (e)$.  At $r< r_{\rm
i}(e)$, the function $a^2_{{\rm z}-}(r;  e)$ grows from its zero point
up  to $r=r_{\rm i}(e)$,  it is  physically irrelevant  up to  $r= 1$,
where  $a^2_{{\rm z}-}(r=1,  e) =  1- e^2$.   The  function $a^2_{{\rm
c}+}(r; e)$  diverges at  $r=e^2$; it has  a local  minimum $a^2_{{\rm
c}+{\rm (min)}}(e)$ at $r_{{\rm z}({\rm min})}(e)$ and a local maximum
$a^2_{{\rm  c}+({\rm  max})}(e)$ at  $r_{{\rm  c}({\rm max})}(e)$;  of
course $r_{{\rm z}({\rm min})} <  r_{{\rm z}({\rm max})} < r_{\rm i}$.
Relating $a^2_{{\rm z}+({\rm min})}(e)$ to $a^2 = 1-e^2$ enables us to
give   the   classification   of  black-hole   and   naked-singularity
backgrounds according to the number of turning points of the embedding
diagrams.  (A classification of  these backgrounds relating the number
of turning points and the number of embeddable regions can be realized
by  relating $a^2_{e4({\rm min})}$  with $a^2_{{\rm  c}+({\rm max})}$,
and $a^2_{e3({\rm max},2)}$ with $a^2_{{\rm z}+({\rm min})}$.)

The subclassification according to the number of turning points can be
given in the following way (with $a^2$ growing):

\begin{center}
\begin{tabular}{llll}
\hline
Subclass & Interval of $e^2$      & Black holes & Naked singularities\\ \hline 
Ta$_1$   & $\bra0,0.18275)$       & $1,3$       & $4,2$              \\
Ta$_2$   & $\bra0.18275,0.51033)$ & 1           & $2,4,2$            \\ \hline
\end{tabular}
\end{center}
Examples   of   subclasses   Ta$_1$    and   Ta$_2$   are   given   in
Figs~\ref{r-a2-016}    and     \ref{r-a2-019}    or    \ref{r-a2-025},
respectively.

\subsubsection{Class Tb: {\protect\boldmath$e^2\in (0.51033,1.125)$}}

If  $e^2$ equals  to  the  critical value  $e^2_{\rm  cr} =  0.51033$,
$r_{\rm  i}(e) =  r_{\rm o}(e)$,  and the  two branches  of $a^2_{{\rm
c}+}$ and  $a^2_{{\rm z}-}$  coalesce. For $e^2  > e^2_{\rm  cr}$, the
reality condition (\ref{e28})  is satisfied everywhere, and $a^2_{{\rm
c}-}(r; e)$, and $a^2_{{\rm z}+}(r; e)$ give two separated branches of
the turning  points. The  function $a^2_{{\rm z}-}(r;  e)$ determining
the  lower branch  is physically  irrelevant at  $r<1$,  being located
between the horizons,  and it has a local  maximum $a^2_{{\rm z}-({\rm
max})}(e)$ at  $r>1$. The function  $a^2_{{\rm z}+}(r; e)$  giving the
upper branch diverges  at $r=e^2$ and for $r  \rightarrow \infty$, and
it can have a local  minimum $a^2_{{\rm z}+({\rm min})}$ near $r=e^2$,
or two local minima $a^2_{{\rm z}+({\rm min},1)}$, $a^2_{{\rm z}+({\rm
min},2)}$  and a  local  maximum $a^2_{{\rm  c}+({\rm max})}$  between
them. Now,  the subclassification according  to the number  of turning
points can be given as follows (with $a^2$ growing):

\begin{center}
\begin{tabular}{llll}
\hline
Subclass & Interval of $e^2$      & Black holes & Naked singularities\\ \hline 
Tb$_1$   & $\bra0.51033,0.52480)$ & 1           & $2,0,2,4,2$        \\
Tb$_2$   & $\bra0.52480,1)$       & 1           & $2,0,2$            \\
Tb$_3$   & $\bra1,1.125)$         & none        & $2,0,2$            \\ \hline
\end{tabular}
\end{center}
An example of subclass Tb$_2$ is given in Fig.\,\ref{r-a2-090}.

\subsubsection{Class Tc: {\protect\boldmath$e^2\in (1.125,\infty)$}}

Now, the function $a^2_{{\rm  c}-}(r; e)$ is irrelevant, being shifted
to  negative values  completely. The  function $a^2_{{\rm  c}+}(r; e)$
behaves  in  the  same  way  as  that  of  the  subclass  Tb$_3$.  The
classification  according  to  the  number  of turning  point  is  the
following:

\begin{center}
\begin{tabular}{llll}
\hline
Subclass & Interval of $e^2$   & Black holes & Naked singularities\\ \hline 
Tc$_1$   & $\bra1.125,\infty)$ & none        & $0,2$              \\ \hline
\end{tabular}
\end{center}

We can  conclude that outside  the outer black-hole horizon,  there is
always  an embeddable  region,  covering exterior  of  the black  hole
except  a  small  part in  the  vicinity  of  the outer  horizon,  and
containing just  one turning  point corresponding to  a throat  of the
embedding diagram. So, the situation is the same as for Schwarzschild,
Reissner\nd  Nordstr\"om, and  Kerr black  holes. On  the  other hand,
under  the inner horizon,  the embeddable  region contains  no turning
point, or two turning points corresponding to a throat and a belly; we
do not consider the case of coalescing of the two turning points in an
inflex point separately.

For  the naked-singularity  backgrounds,  the situation  is much  more
complex,  including  various  possibilities   of  the  number  of  the
embeddable regions and the turning points of the diagrams. There are a
lot  of  cases  that  cannot  appear  with  Kerr  naked  singularities
\cite{SH99c}. As examples, let us mention one region with four turning
points, and  one region with  no turning point.  Again,  in situations
which  are  necessary  in  order  to determine  behaviour  of  typical
embedding  diagrams,  the  curves  $a^2_{{\rm z}\pm}(r;e)$  are  given
explicitly (see Figs~\ref{r-a2-016}--\ref{r-a2-090}).

\subsection{Embeddability of photon circular orbits}

The  motion of  a photon  in the  equatorial plane  of  Kerr\nd Newman
spacetimes is  determined by the function (see,  e.g., \cite{BBS})
\be
  R(r;a,e,E,\Phi) = [Er^2 - a(\Phi -  aE)]^2 - \Delta (\Phi - aE)^2,
\ee
where $E$  is the covariant  energy of the  photon, and $\Phi$  is its
axial angular momentum.  For photon circular orbits the conditions
\be
  R = 0, \quad \frac{\partial R}{\partial r} = 0           \label{e43}
\ee
must  be satisfied  simultaneously.  The  equatorial photon  motion is
fully governed by the impact parameter
\be
  \ell = \frac{\Phi}{E}.
\ee
It follows  from the conditions  (\ref{e43}) that the radii  of photon
circular orbits are determined by the equation
\be
  r^2 -  3r + 2a^2 + 2e^2 \pm 2a  \Delta^{1/2} = 0;
\ee
the corresponding impact parameter is given by the relation
\be
  \ell = -\frac{a(r^2 +  3r - 2e^2)}{r^2 - 3r + 2e^2}.
\ee
Equivalently, the radii of photon circular orbits can be determined by
\be
  a^2 = a^2_{\rm ph}(r;e) =
    \frac{(r^2 - 3r +  2e^2)^2}{4(r - e^2)}.               \label{e47}
\ee
Clearly, the circular orbits must be located at
\be
  r  \geq e^2.
\ee
The zero  points of
$a^2_{\rm   ph}(r;e)$   are   located    at   radii
\be
  r_1(e) = {\textstyle\frac{1}{2}}[3-(9-8e^2)^{1/2}],            \quad
  r_2(e) = {\textstyle\frac{1}{2}}[3+(9-8e^2)^{1/2}],
\ee
giving   photon    circular   orbits   of    Reissner\nd   Nordstr\"om
spacetimes. The extrema of $a^2_{\rm ph}(r;e)$ are located at the zero
points  $r_1(e)$, $r_2(e)$  (if $e^2  <  \textstyle{\frac{9}{8}}$), at
$r=1$ (if $e^2 < 1$) and at
\be
  r = {\textstyle\frac{4}{3}}e^2,
\ee
where  a   minimum  exists   for  $e^2<\frac{3}{4}$,  a   maximum  for
$\frac{3}{4}<e^2<\frac{9}{8}$, and a minimum for $e^2\geq\frac{9}{8}$.
At $r=\frac{4}{3}e^2$,  the value of  $a^2_{\rm ph}(r;e)$ is  given by
the function
\be
  a^2_{\rm ex}(e) =
    {\textstyle\frac{1}{27}}e^2  (8e^2 - 9)^2.             \label{e52}
\ee

The  function $a^2_{\rm  ex}(e)$ determines  boundary  between Kerr\nd
Newman  spacetimes  containing different  numbers  of photon  circular
orbits. If $e^2 < \frac{3}{4}$, there can be 2 or 4 circular orbits in
the   black-hole  spacetimes   and  2   orbits   in  naked-singularity
spacetimes.  If  $\frac{3}{4}<e^2<1$, there  are 2 circular  orbits in
black-hole and 2 or 4  orbits in naked-singularity spacetimes. For the
naked-singularity spacetimes with  $1<e^2<\frac{9}{8}$, there are 2 or
4 orbits; if $e^2>\frac{9}{8}$, there are 0 or 2 orbits.

\begin{figure}[p]
\def\pcls{0.677}
\centering\leavevmode
\begin{minipage}{.48\hsize}
\centering\leavevmode
{\footnotesize Class Pa: $e^2=0.16$}\par\vskip 1mm\par
\centering\leavevmode\epsfxsize=\pcls\hsize\epsfbox{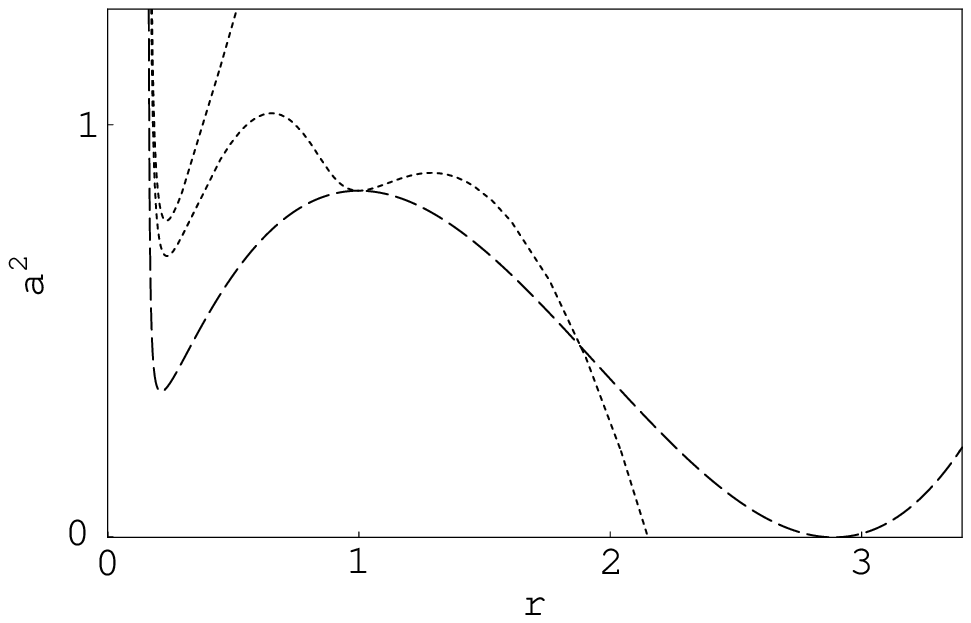}
\par\centering\leavevmode
{\footnotesize Class Pc: $e^2=0.78$}\par\vskip 1mm\par
\centering\leavevmode\epsfxsize=\pcls\hsize\epsfbox{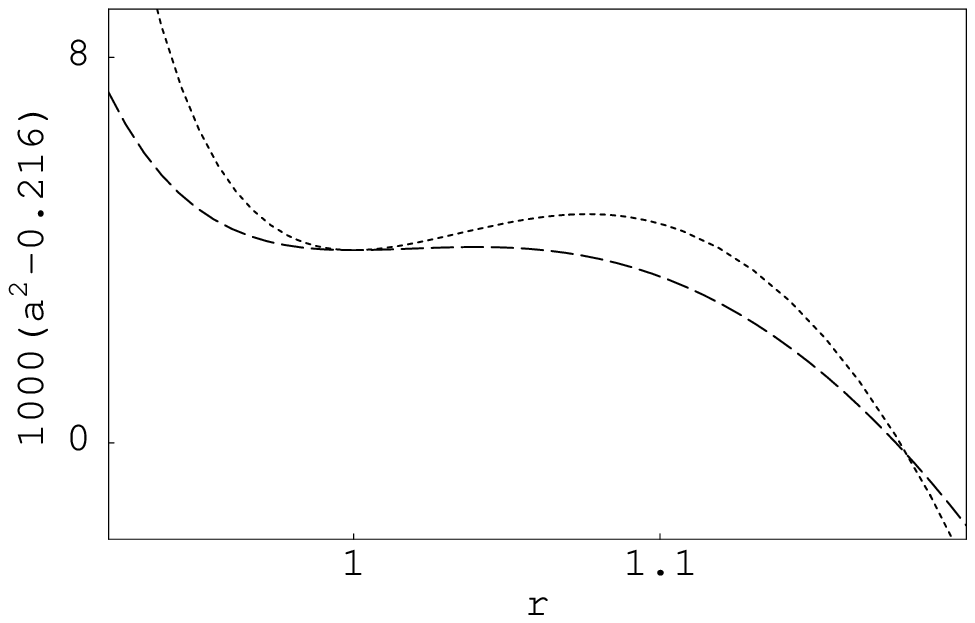}
\par\centering\leavevmode
{\footnotesize Class Pe: $e^2=0.81$}\par\vskip 1mm\par
\centering\leavevmode\epsfxsize=\pcls\hsize\epsfbox{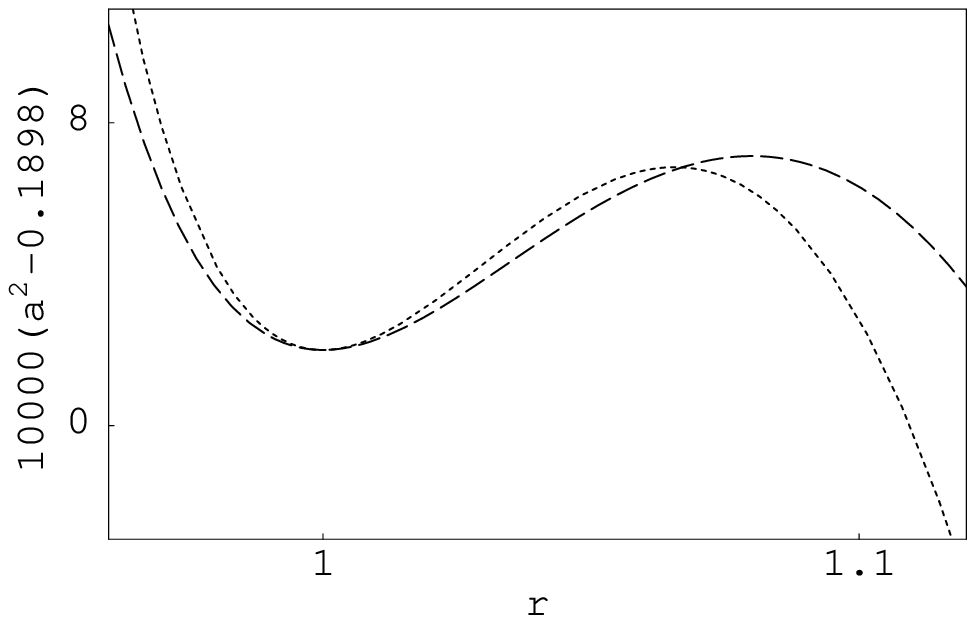}
\par\centering\leavevmode
{\footnotesize Class Pg: $e^2=0.86$}\par\vskip 1mm\par
\centering\leavevmode\epsfxsize=\pcls\hsize\epsfbox{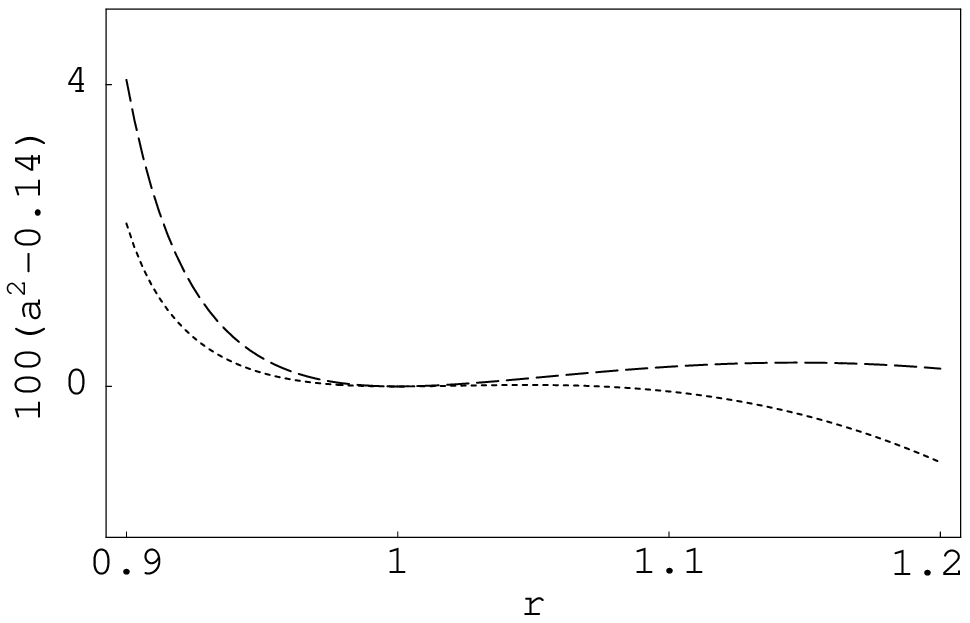}
\par\centering\leavevmode
{\footnotesize Class Pi: $e^2=1.3$}\par\vskip 1mm\par
\centering\leavevmode\epsfxsize=\pcls\hsize\epsfbox{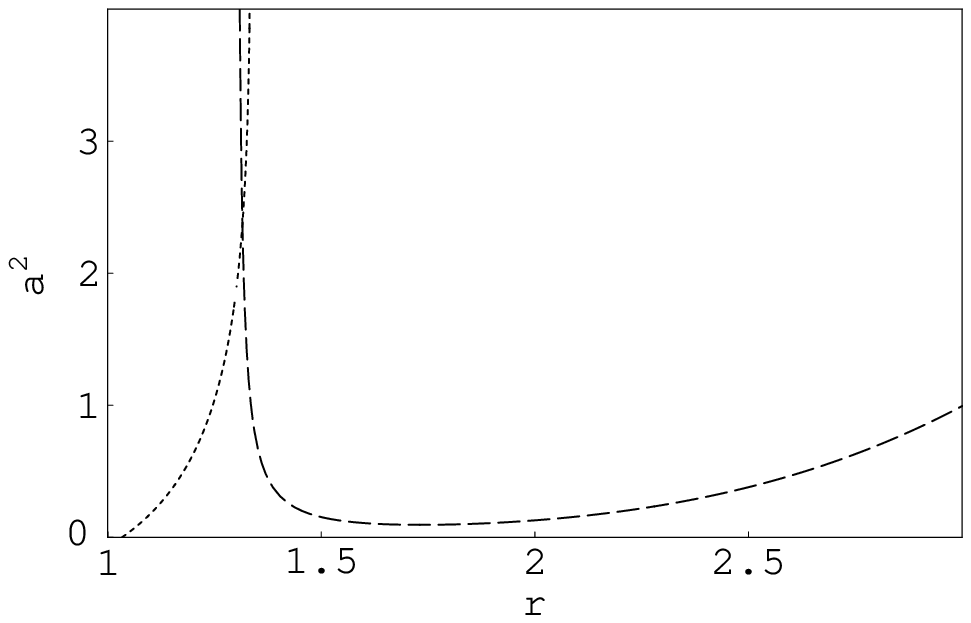}
\end{minipage}\hfill%
\begin{minipage}{.48\hsize}
\centering\leavevmode
{\footnotesize Class Pb: $e^2=0.35$}\par\vskip 1mm\par
\centering\leavevmode\epsfxsize=\pcls\hsize\epsfbox{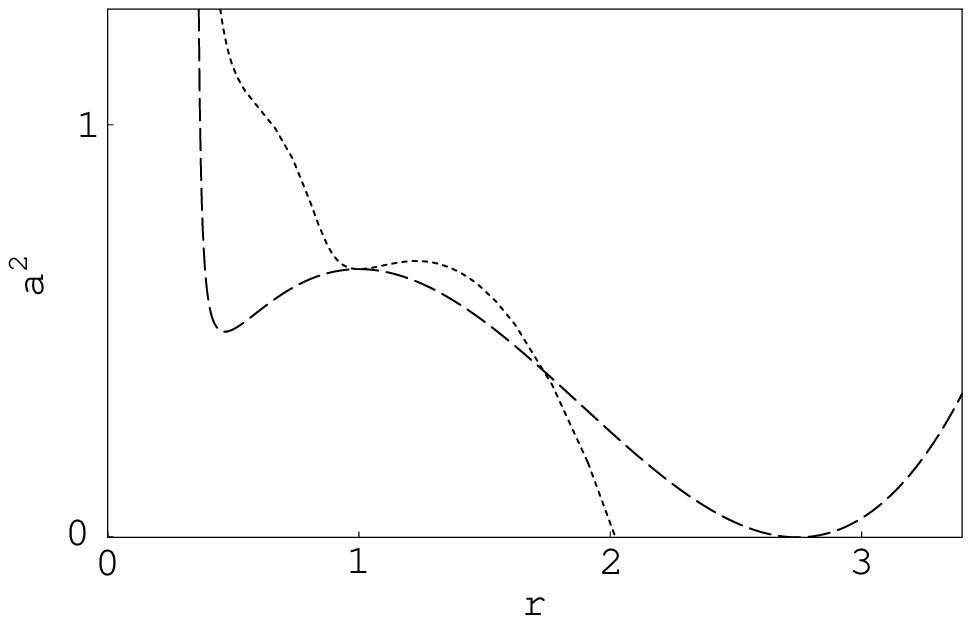}
\par\centering\leavevmode
{\footnotesize Class Pd: $e^2=0.804$}\par\vskip 1mm\par
\centering\leavevmode\epsfxsize=\pcls\hsize\epsfbox{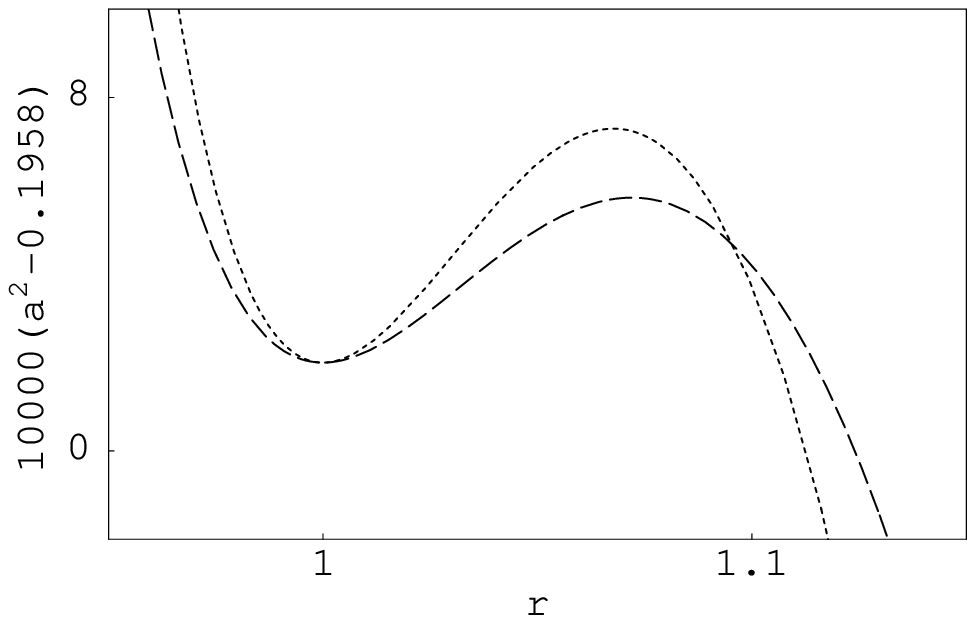}
\par\centering\leavevmode
{\footnotesize Class Pf: $e^2=0.822$}\par\vskip 1mm\par
\centering\leavevmode\epsfxsize=\pcls\hsize\epsfbox{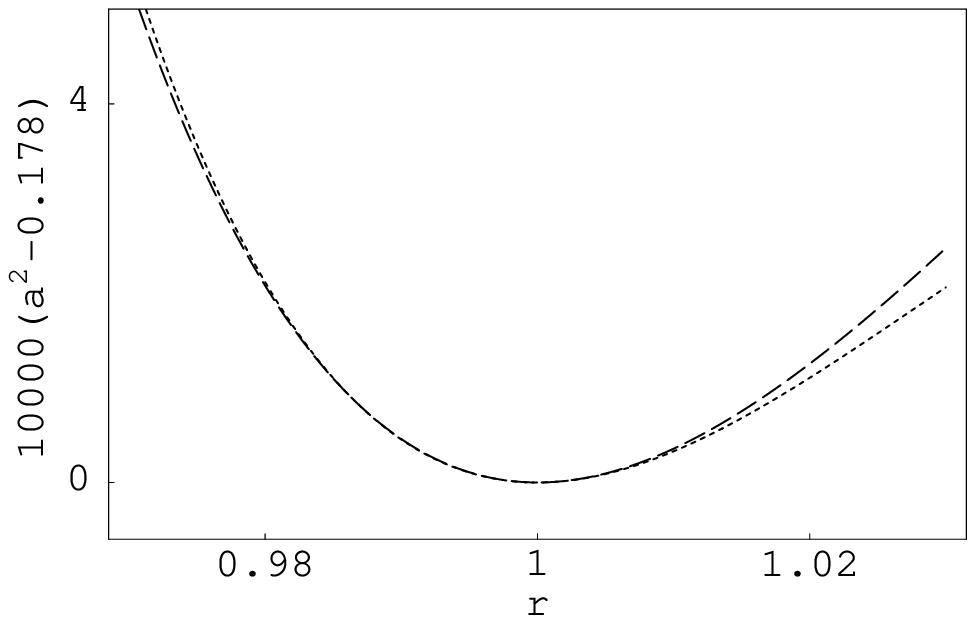}
\par\centering\leavevmode
{\footnotesize Class Ph: $e^2=1.05$}\par\vskip 1mm\par
\centering\leavevmode\epsfxsize=\pcls\hsize\epsfbox{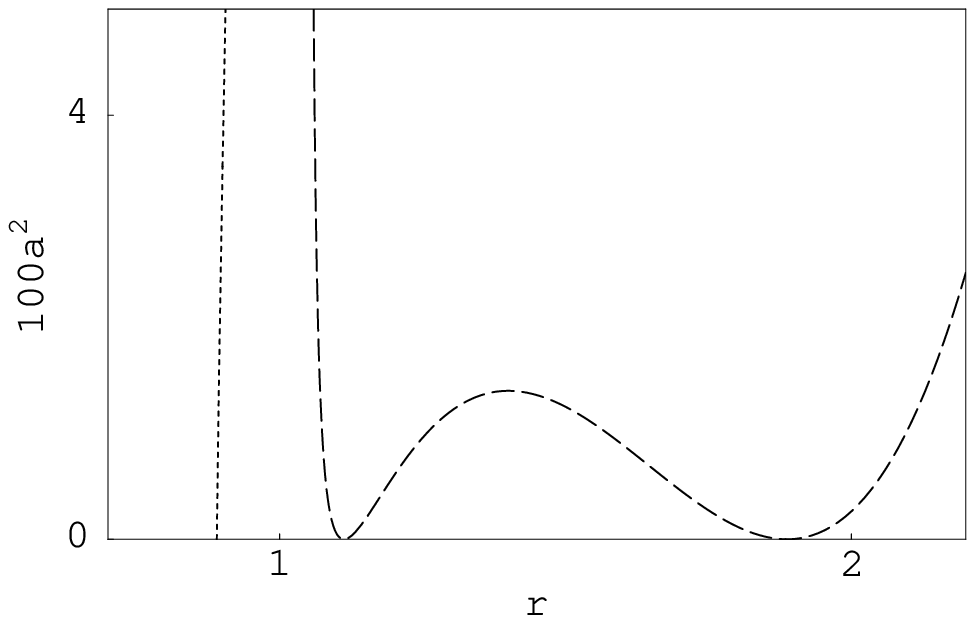}
\par\centering\leavevmode
{\footnotesize Class Pj: $e^2=2.3$}\par\vskip 1mm\par
\centering\leavevmode\epsfxsize=\pcls\hsize\epsfbox{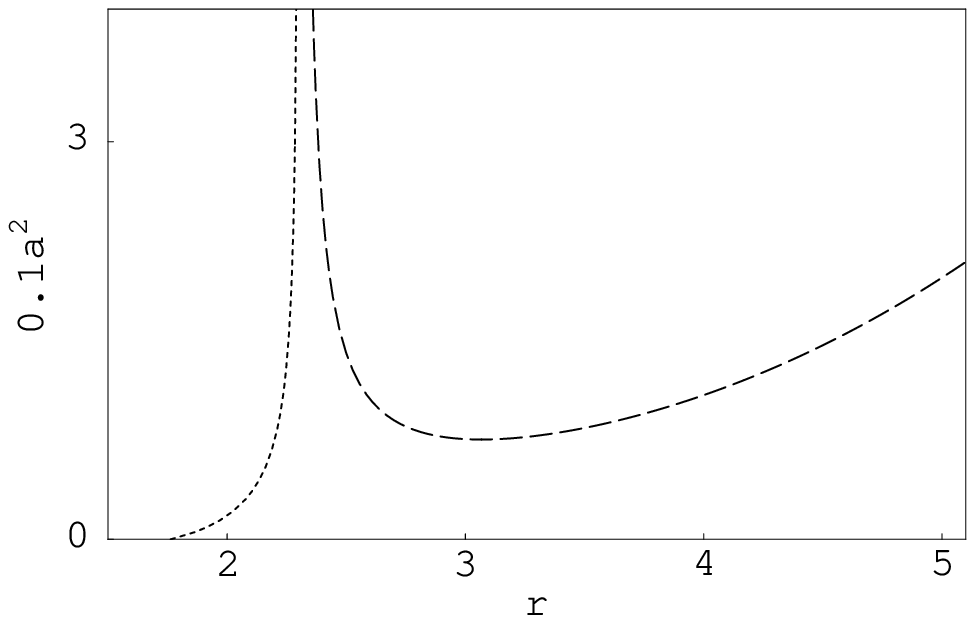}
\end{minipage}
\caption{Embeddability of  photon  circular  orbits in  the  embedding
  diagrams of the optical geometry.  It is determined by the relations
  of the curves $a^2_{\rm e}(r;e)$ (dotted lines) giving the limits of
  embeddability, and  $a^2_{\rm ph}(r;e)$ (dashed  lines) giving radii
  of  the photon  circular  orbits.   We present  all  of the  typical
  situations   representing  the   classes  Pa--Pj,   restricting  our
  attention on the relevant  parts of the functions $a^2_{\rm e}(r;e)$
  and  $a^2_{\rm ph}(r;e)$. Note  that the  circular orbits  under the
  inner black-hole horizon are always outside the embeddable regions.}
\label{r-a2cpe}
\end{figure}

\begin{figure}[t]
\centering\leavevmode \epsfxsize=\hsize \epsfbox{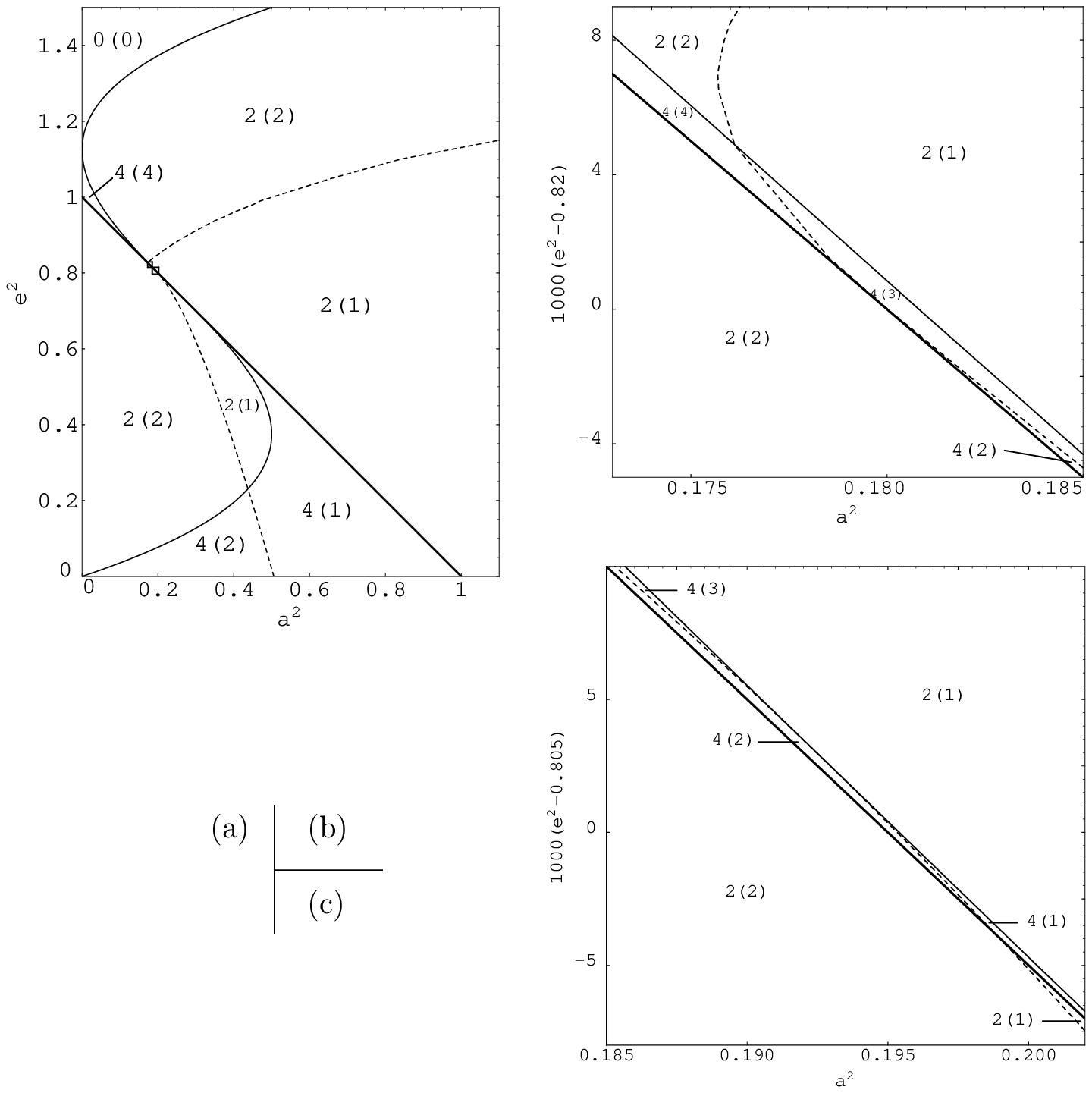}
\caption{Classification of the Kerr\nd Newman  spacetimes according to
  em\-bed\-dability of photon circular orbits. The solid lines separate
  regions of the parameter $a^2$-$e^2$ corresponding to spacetimes containing
  different numbers of the circular orbits; the bold solid straight line
  correspond to extreme black-hole spacetimes.  The dashed line separates
  regions of the parameter space with different numbers of the embeddable
  circular orbits. The S-shaped solid curve is given by
  Eq.~(\protect\ref{e52}). The corresponding region of the parameter space are
  depicted by pairs of digits, the first one expressing number of the circular
  orbits, the second one (in parentheses) the number of the embeddable orbits.
  (a)~the global view of the classification, (b,c)~details of some narrow
  regions located in the vicinity of the extreme black-hole states, where the
  structure is rather complicated.}
\label{a2-e2}
\end{figure}

In the case of extreme  black holes ($a^2+e^2=1$), the radii of photon
circular orbits and corresponding values of the impact parameter are
\be
\begin{array}{ll}                                              
  r = 1,     &  \quad \ell = a+1/a,    \\
  r = 2-2a,  &  \quad \ell = 4-3a,     \\
  r = 2+2a,  &  \quad \ell = -(4+3a),
\end{array}
\ee
The  counterrotating orbit  at $r=2+2a$  is always  located  above the
horizon.   The  corotating orbit  at  $r=2-2a$  is  located above  the
horizon  if  $e^2  >  \frac{3}{4}$,  and under  the  horizon  if  $e^2
<\frac{3}{4}$. There  are two orbits  at $r=1$ if  $e^2 <\frac{3}{4}$,
but no circular orbit at  $r=1$, if $e^2 >\frac{3}{4}$ (see \cite{BBS}
for details).

Now, we shall discuss in  which cases the photon circular orbits enter
the regions of embeddability. Radii  of the photon circular orbits are
determined   by   the    function   $a^2_{\rm   ph}(r;e)$   given   by
Eq.~(\ref{e47}). The embeddability of  these orbits must be determined
by  a  numerical  procedure.    Note  that,  generally,  the  function
$a^2_{{\rm z}-}(r;e)$  has common  points with $a^2_{\rm  ph}(r;e)$ at
its zero points (if $e^2<\frac{9}{8}$), and at $r=1$ (if $e^2 < 1$).

The classification  of the Kerr\nd Newman spacetimes  according to the
embeddability of  the photon circular orbits  can be again  given in a
simple  way. In  the presented  classification scheme  related  to the
parameter $e^2$, the first digit  gives the number of circular orbits,
the  digits  in  parentheses   determine  changes  of  the  number  of
embeddable circular  orbits, with  parameter $a^2$ growing.  Note that
the circular  orbits under  the inner horizon  are always  outside the
embeddable regions:

\begin{center}
\begin{tabular}{llll}
\hline
Class & Interval of $e^2$          & Black holes   & Naked singularities\\
\hline 
Pa    & $\bra0,0.23083)$           & $2(2),4(2,1)$ & 2(1)               \\
Pb    & $\bra0.23083,\frac{3}{4})$ & $2(2,1),4(1)$ & 2(1)               \\
Pc    & $\bra\frac{3}{4},0.80115)$ & $2(2,1)$      &  $4(1),2(1)$       \\
Pd    & $\bra0.80115,0.80797)$     & 2(2)          &  $4(2,1),2(1)$     \\
Pe    & $\bra0.80797,0.82040)$     & 2(2)          &  $4(2,3),2(1)$     \\
Pf    & $\bra0.82040,0.82488)$     & 2(2)          &  $4(4,3),2(1)$     \\
Pg    & $\bra0.82488,1)$           & 2(2)          &  $4(4),2(2,1)$     \\
Ph    & $\bra1,\frac{9}{8})$       & none          &  $4(4),2(2,1)$     \\
Pi    & $\bra\frac{9}{8},1.70233)$ & none          &  $0(0),2(2,1)$     \\
Pj    & $\bra1.70233,\infty)$      & none          &  $0(0),2(2)$       \\
\hline
\end{tabular}
\end{center}

The embeddability of the photon circular orbits can be easily read out
from  the sequence of  figures (Fig.\,\ref{r-a2cpe})  representing the
classification given above.

Details  of the  classification, i.e.,  distribution in  the parameter
space  of the  Kerr\nd Newman  backgrounds,  must be  determined by  a
numerical procedure.  The  results are presented in Fig.\,\ref{a2-e2}.
The  regions of  the parameter  plane $a^2$-$e^2$  are denoted  by two
digits.  The first  one gives the number of  photon circular orbits in
the corresponding  background, the  second one (in  parentheses) gives
the number of embeddable orbits.

\subsection{Construction of the embedding diagrams}

The relevant  properties of the  embedding diagrams are  determined by
the  functions  $a^2_{\rm e}(r;e)$  and  $a^2_{\rm z}(r;e)$  governing
embeddable parts of the  optical reference geometry and turning points
of the  diagrams.  We  shall present a  classification of  the Kerr\nd
Newman spacetimes  according to the number of  the embeddable regions,
and  the   number  of   their  turning  points.   All  cases   of  the
classification   will   be   represented   by  a   typical   embedding
diagram.  Inspecting all  of the  types of  the behaviour  of functions
$a^2_{{\rm  c}\pm}(r;e)$   (given  by  classes  Ta\nd   Tc  and  their
subclasses), we  find that there  are 4 cases  of the behaviour  of the
embeddings  for  the  black-hole  spacetimes,  and 11  cases  for  the
naked-singularity  spacetimes. A complete  list of  typical embeddings
will  be  given by  using  behaviour  of  the characteristic  functions
$a^2_{\rm e}(r;e)$  and $a^2_{\rm z}(r;e)$ for some  typical values of
the  parameter $e^2$---see  Figs~\ref{r-a2-016}--\ref{r-a2-090}, where
the functions $a^2_{\rm h}(r;e)$  and $a^2_{\rm ph}(r;e)$ are included
for  completeness. In  order to  obtain  all of  the 15  types of  the
embedding diagrams, it  is necessary to consider at  least four values
of the parameter $e^2$ belonging subsequently to the subclasses Ea$_1$
($e^2=0.16$),   Ea$_3$  ($e^2=0.19$),   Ea$_6$   ($e^2=0.25$),  Eb$_1$
($e^2=0.92$)  of   the  classification  according  to   the  number  of
embeddable  regions,  discussed   above.  The  behaviour  of  $a^2_{\rm
c}(r;e)$  for these  four  values of  $e^2$  enables us  to present  a
complete  list of  typical  embedding diagrams.  The  classes will  be
denoted  successively  with growing  parameter  $a^2$  for each  fixed
$e^2$.

In  the  regions  of  the  optical geometry  where  the  embeddability
condition $E(r;a,e) \geq 0$ is satisfied, the embedding diagram can be
constructed   for   a  fixed   parameter   $a$   by  integrating   the
parametrically    expressed    embedding    formula    $z(r)$    (cf.\
Eq.~(\ref{e37})), and transferring it into the final form $z(\rho)$ by
an  appropriate  numerical procedure  using  (\ref{e35}).  In all  the
typical diagrams also photon  circular orbits are illustrated, if they
enter the  embeddable regions. (Of  course, the presented  diagrams do
not cover  all of  the possibilities for  the embeddability  of photon
circular orbits.)

\begin{figure}[p]
\begin{minipage}[c]{.36\hsize}
\centering \leavevmode \epsfxsize=.9\hsize \epsfbox{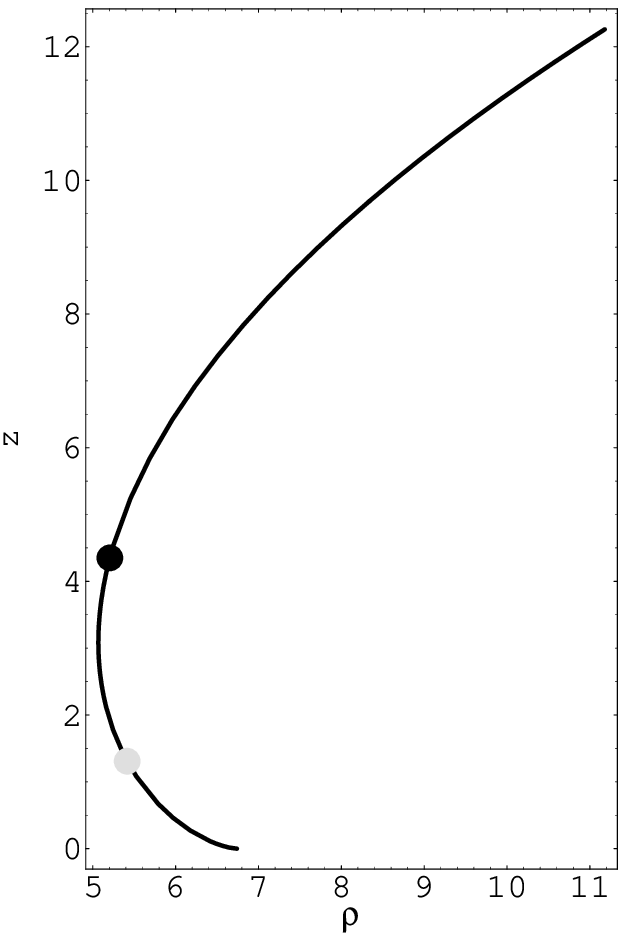}
\end{minipage}\hfill%
\begin{minipage}[c]{.62\hsize}
\centering \leavevmode \epsfxsize=.85\hsize \epsfbox{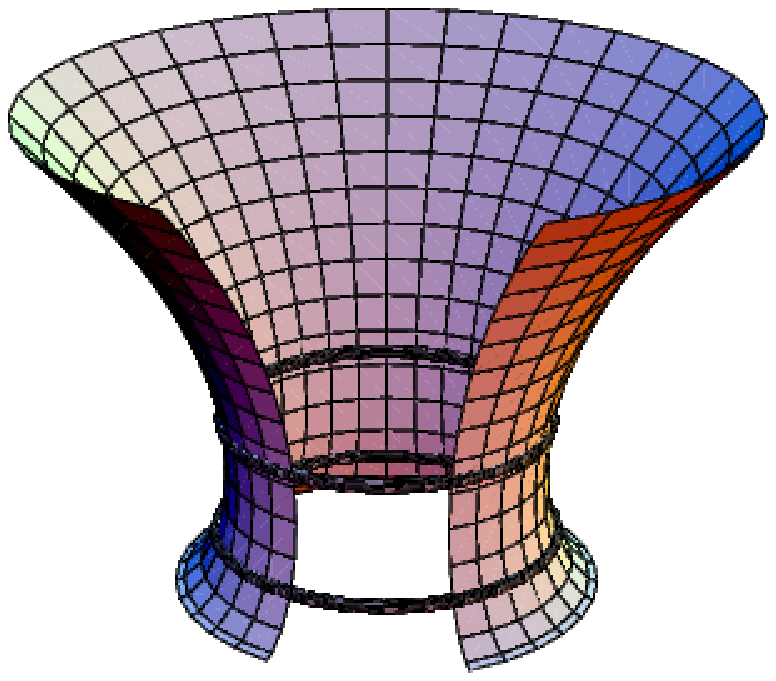}
\end{minipage}
\caption{Embedding diagram of the  Kerr\nd Newman  black holes  of the
  type BH$_1$,  constructed for  $a^2=0.16$, $e^2=0.16$. The  rings in
  the  3D diagram  represent photon  circular orbits.  Both corotating
  (gray  spot in  2D diagram)  and counterrotating  (black spot  in 2D
  diagram) are  displaced from  the throat of  the diagram,  where the
  centrifugal  force  vanishes. This  is  a  general  property of  the
  rotating backgrounds.}
\label{ed-bh1}
\end{figure}

\begin{figure}[p]
\begin{minipage}[b]{.32\hsize}
\centering\leavevmode \epsfxsize=\hsize \epsfbox{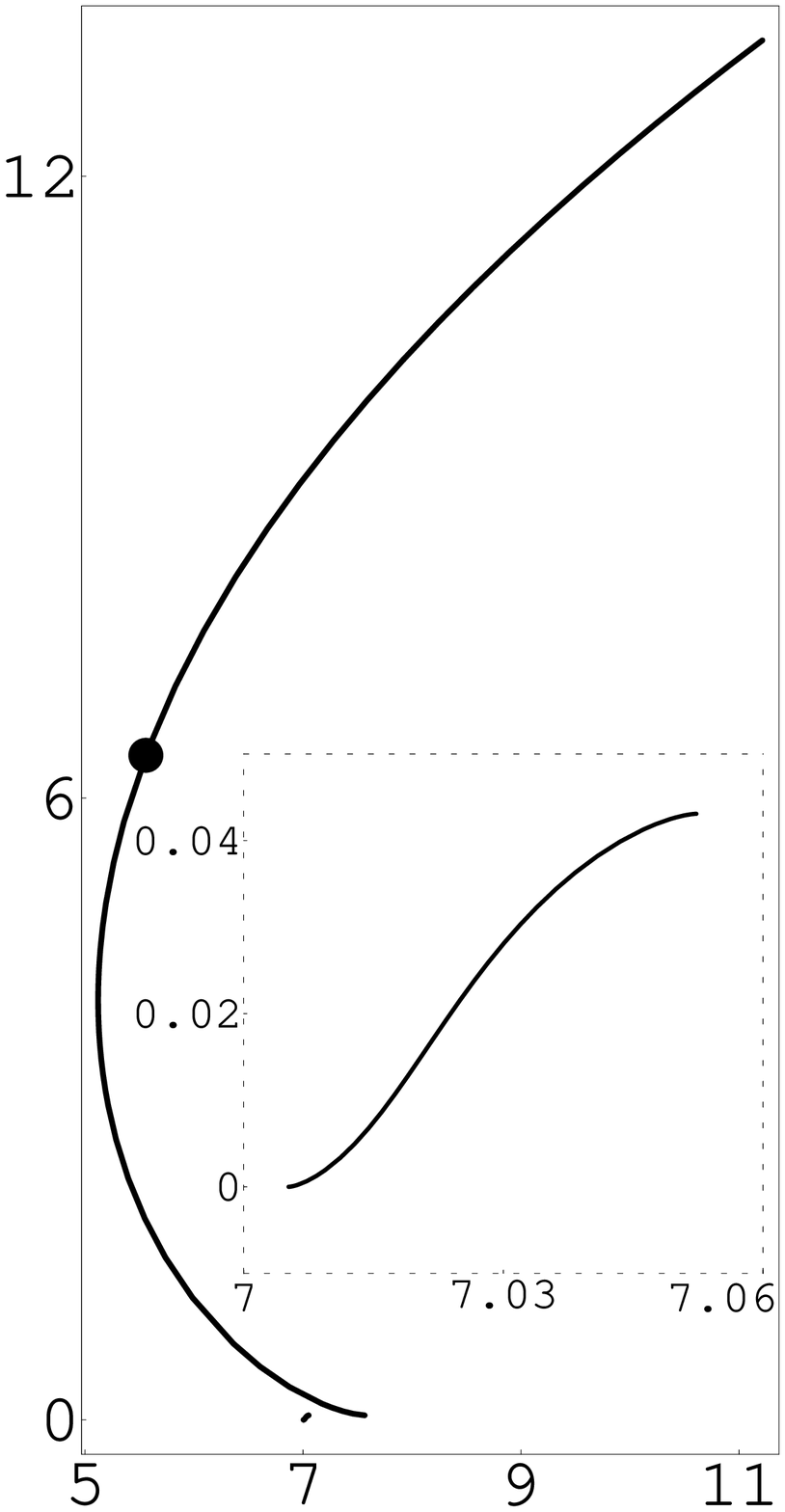}
\par {\small (a)}
\end{minipage}\hfill%
\begin{minipage}[b]{.32\hsize}
\centering\leavevmode \epsfxsize=.962\hsize \epsfbox{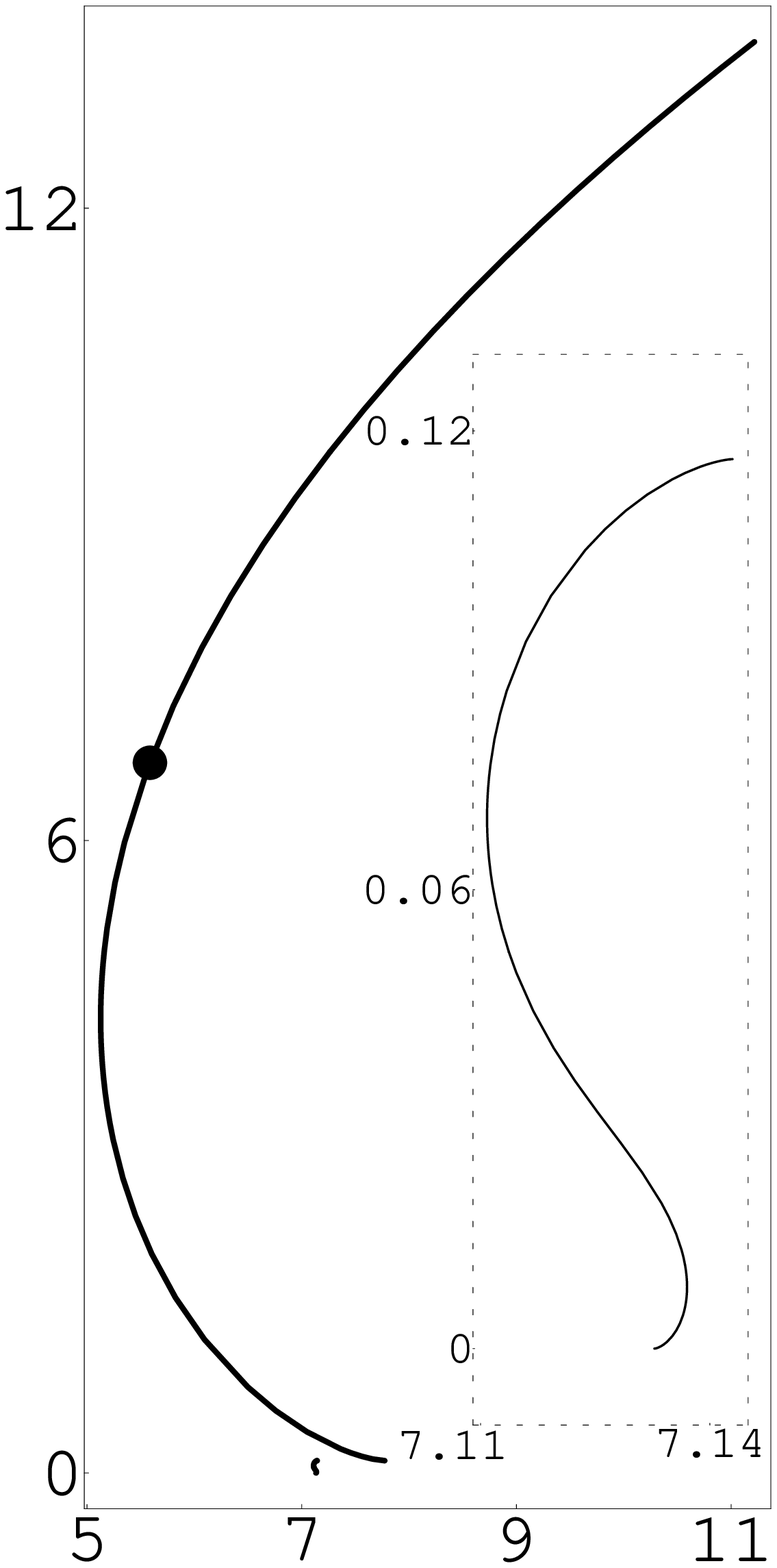}
\par {\small (b)}
\end{minipage}\hfill%
\begin{minipage}[b]{.32\hsize}
\centering\leavevmode \epsfxsize=.932\hsize \epsfbox{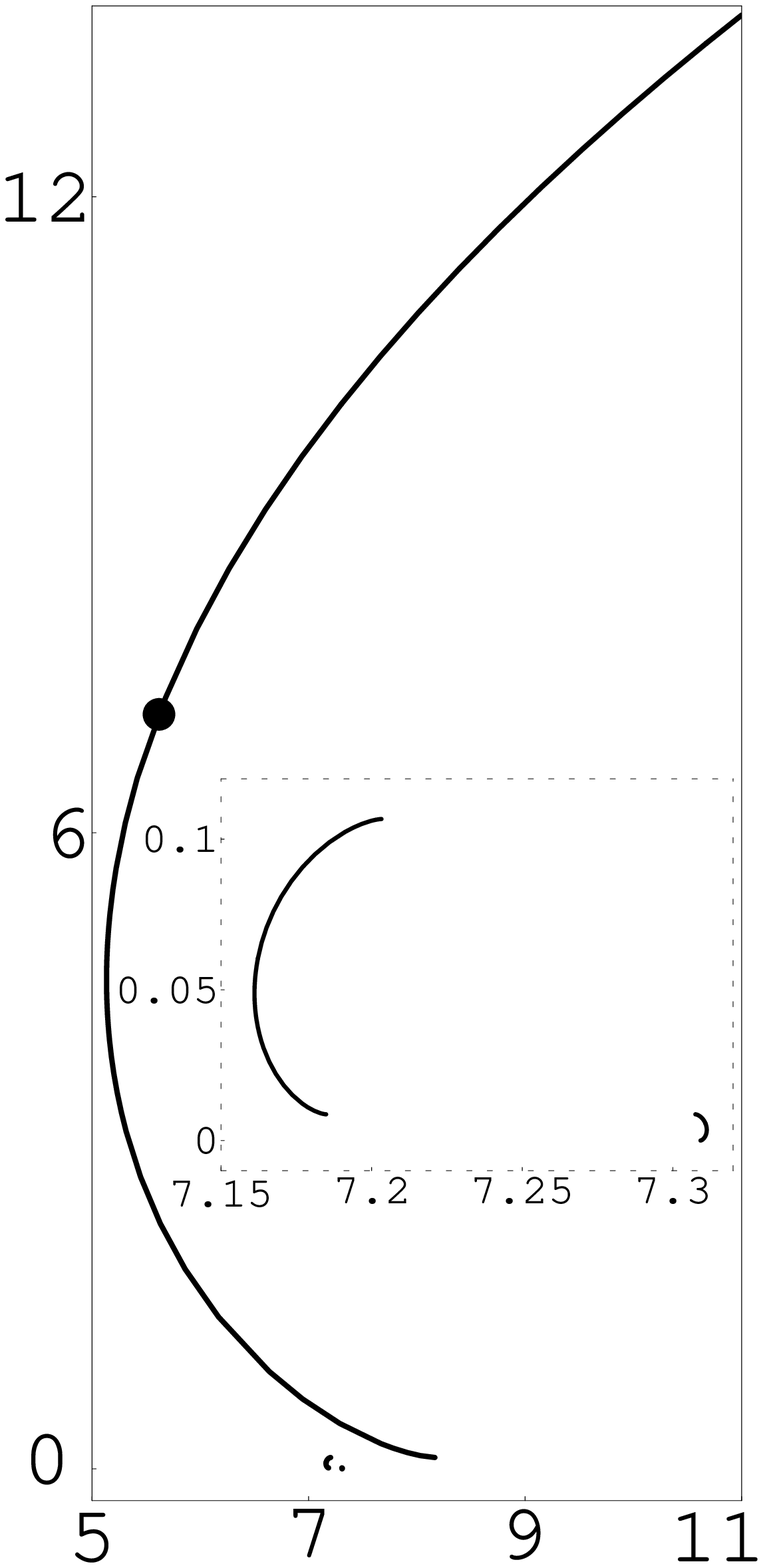}
\par {\small (c)}
\end{minipage}
\caption{Embedding diagrams  of  the  black  holes  classes  BH$_2$\nd
  BH$_4$.  (a)~Class BH$_2$  ($a^2=0.7$, $e^2=0.16$); (b)~class BH$_3$
  ($a^2=0.75$,    $e^2=0.16$);     (c)~class    BH$_4$    ($a^2=0.81$,
  $e^2=0.16$).}
\label{ed-bh2-4}
\end{figure}

\begin{figure}[t]

\begin{minipage}[c]{.4\hsize}
\centering\leavevmode \epsfxsize=\hsize \epsfbox{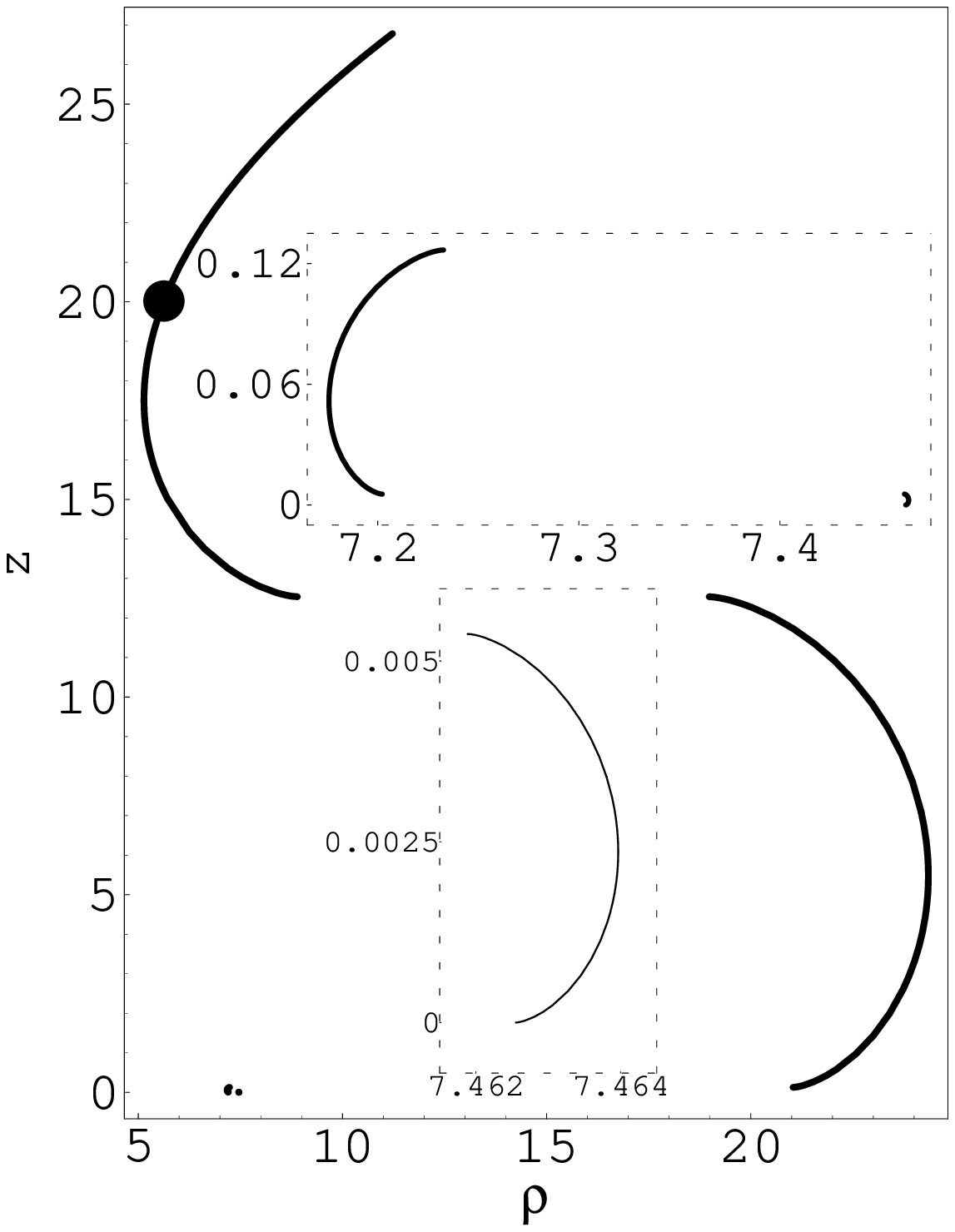}
\end{minipage}\hfill%
\begin{minipage}[c]{.5\hsize}
\centering\leavevmode \epsfxsize=\hsize \epsfbox{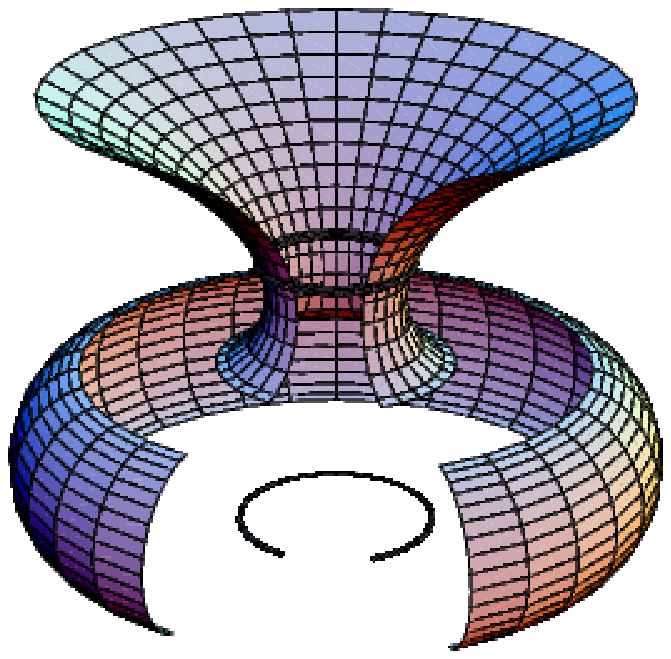}
\end{minipage}

\caption{Embedding diagram   of  the  NS$_1$   type,  constructed  for
  $a^2=0.86$, $e^2=0.16$.}
\label{ed-ns1}
\end{figure}

\begin{figure}[t]
\begin{minipage}[b]{.164\hsize}
\centering\leavevmode \epsfxsize=\hsize \epsfbox{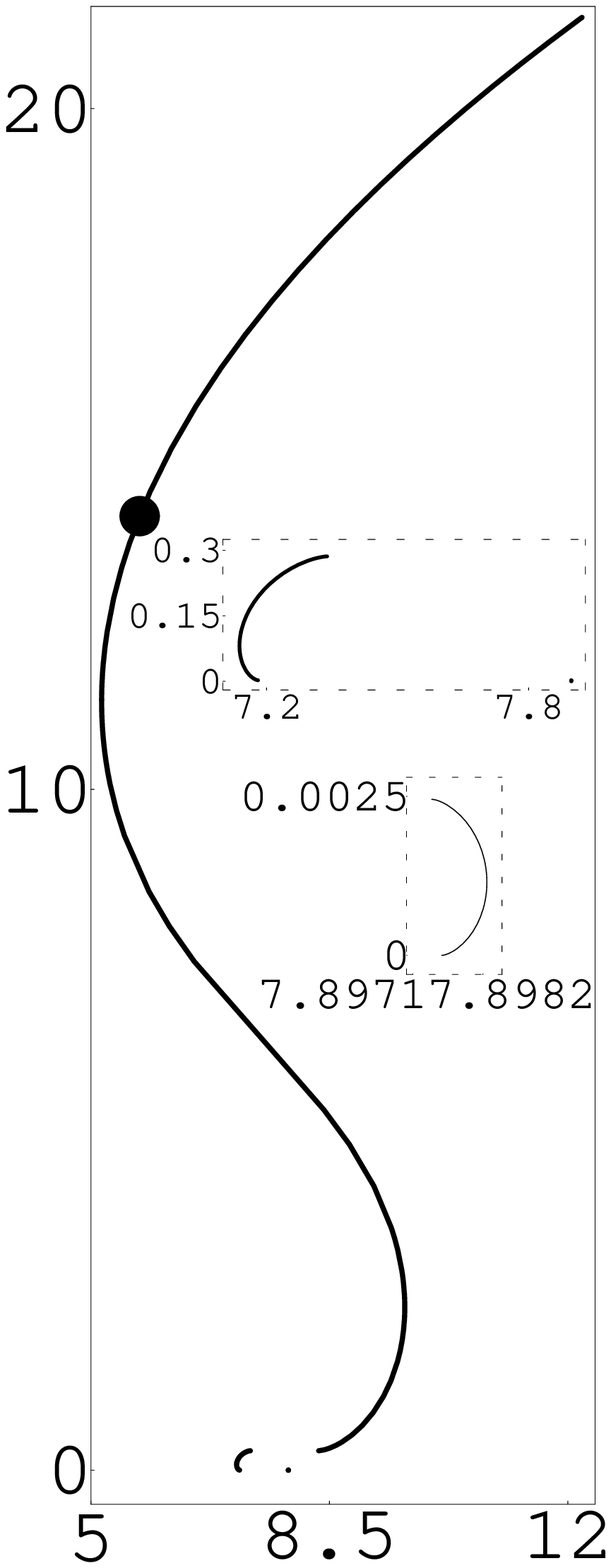}
\par {\small (a)}
\end{minipage}\hfill%
\begin{minipage}[b]{.156\hsize}
\centering\leavevmode \epsfxsize=\hsize \epsfbox{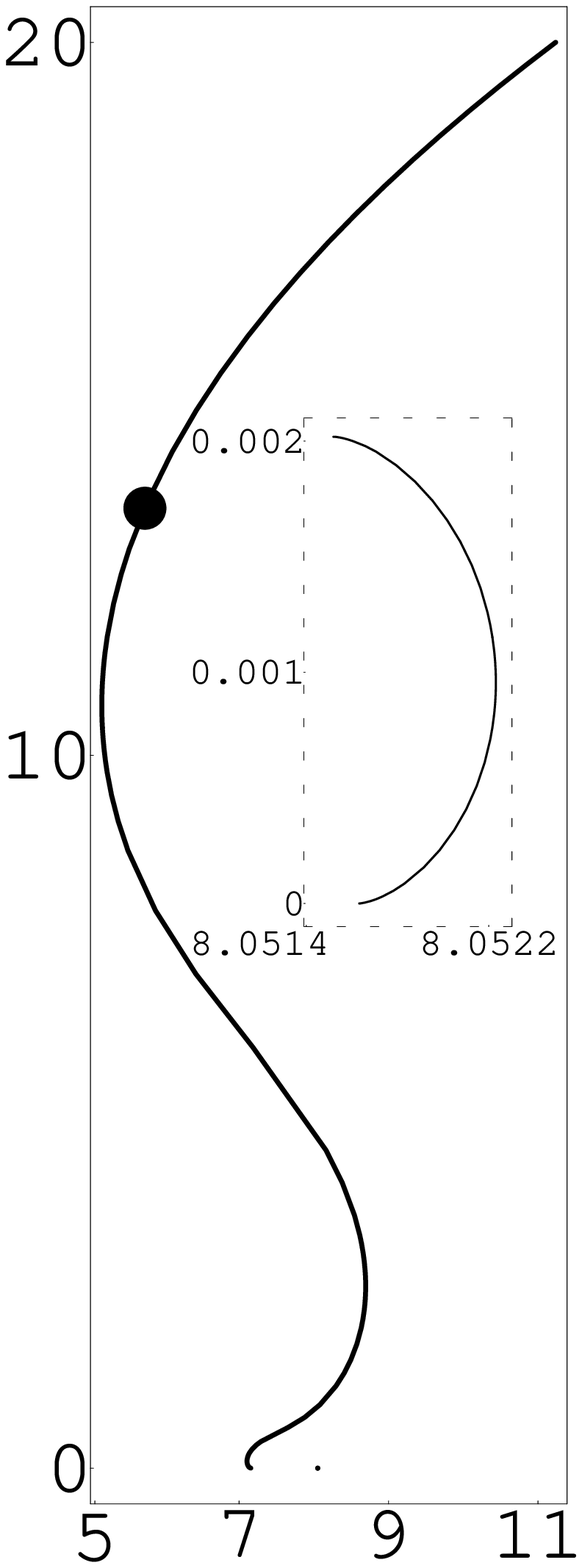}
\par {\small (b)}
\end{minipage}\hfill%
\begin{minipage}[b]{.275\hsize}
\centering\leavevmode \epsfxsize=\hsize \epsfbox{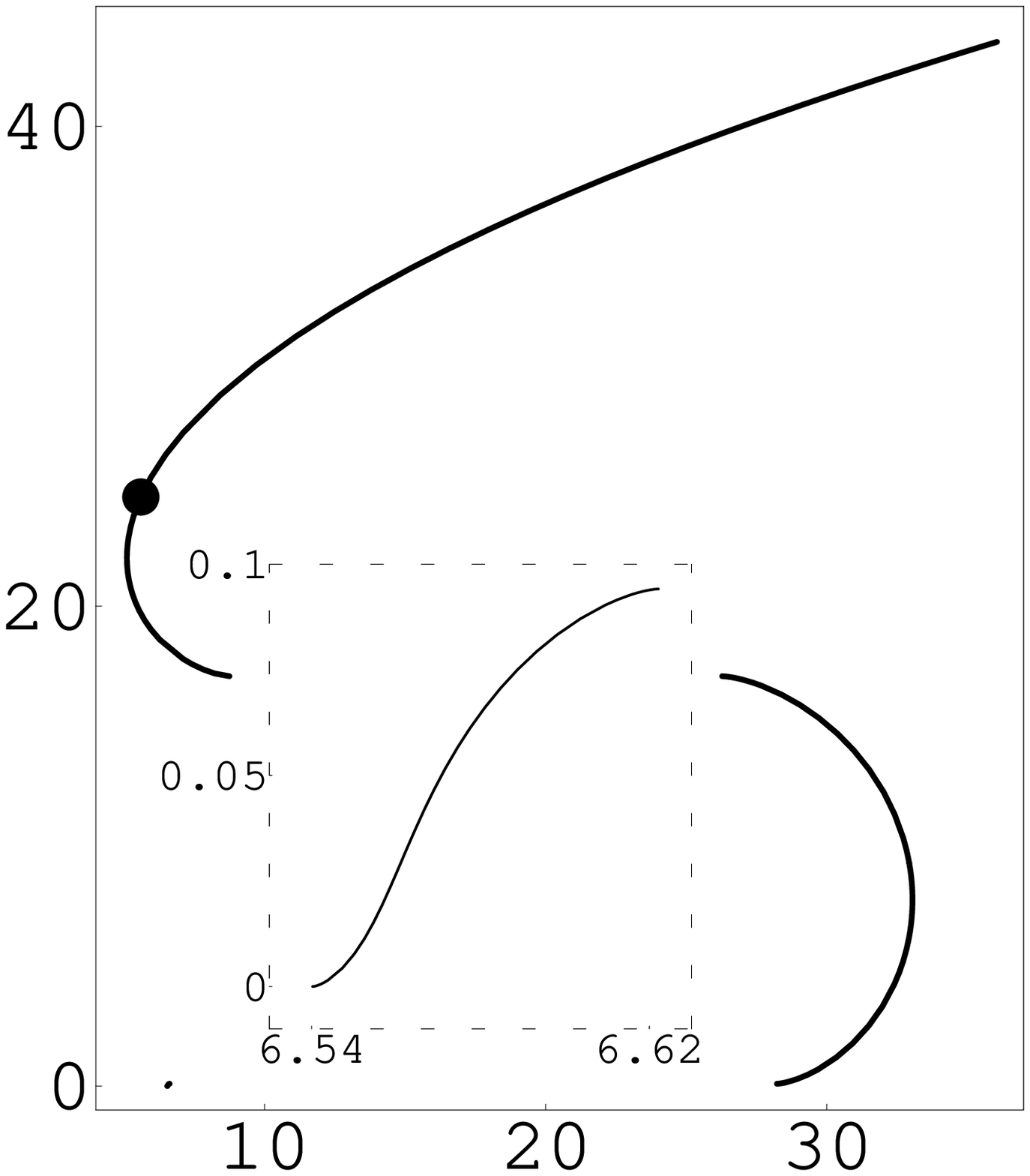}
\par {\small (c)}
\end{minipage}\hfill%
\begin{minipage}[b]{.226\hsize}
\centering\leavevmode \epsfxsize=\hsize \epsfbox{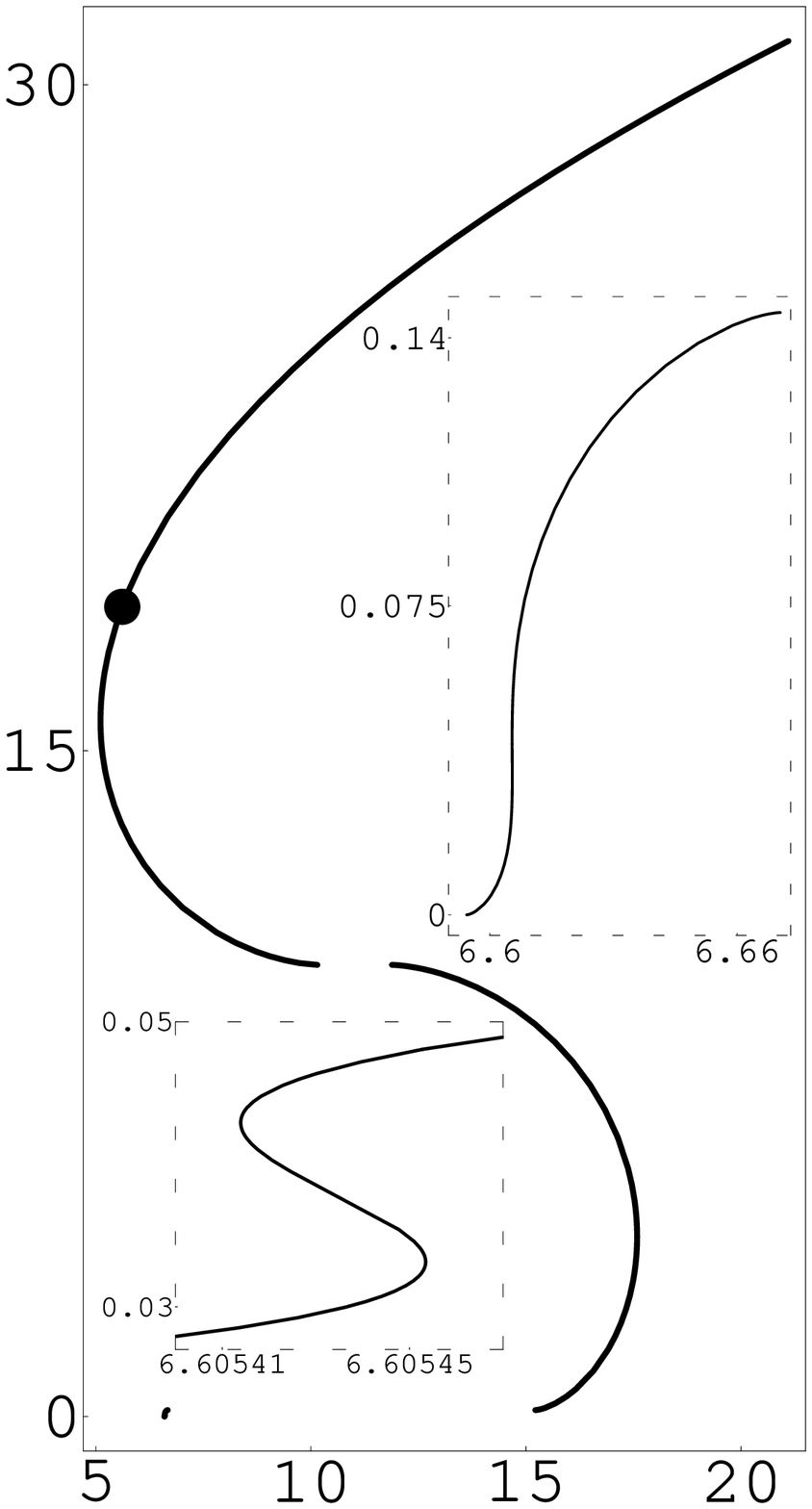}
\par {\small (d)}
\end{minipage}\hfill%
\begin{minipage}[b]{.164\hsize}
\centering\leavevmode \epsfxsize=\hsize \epsfbox{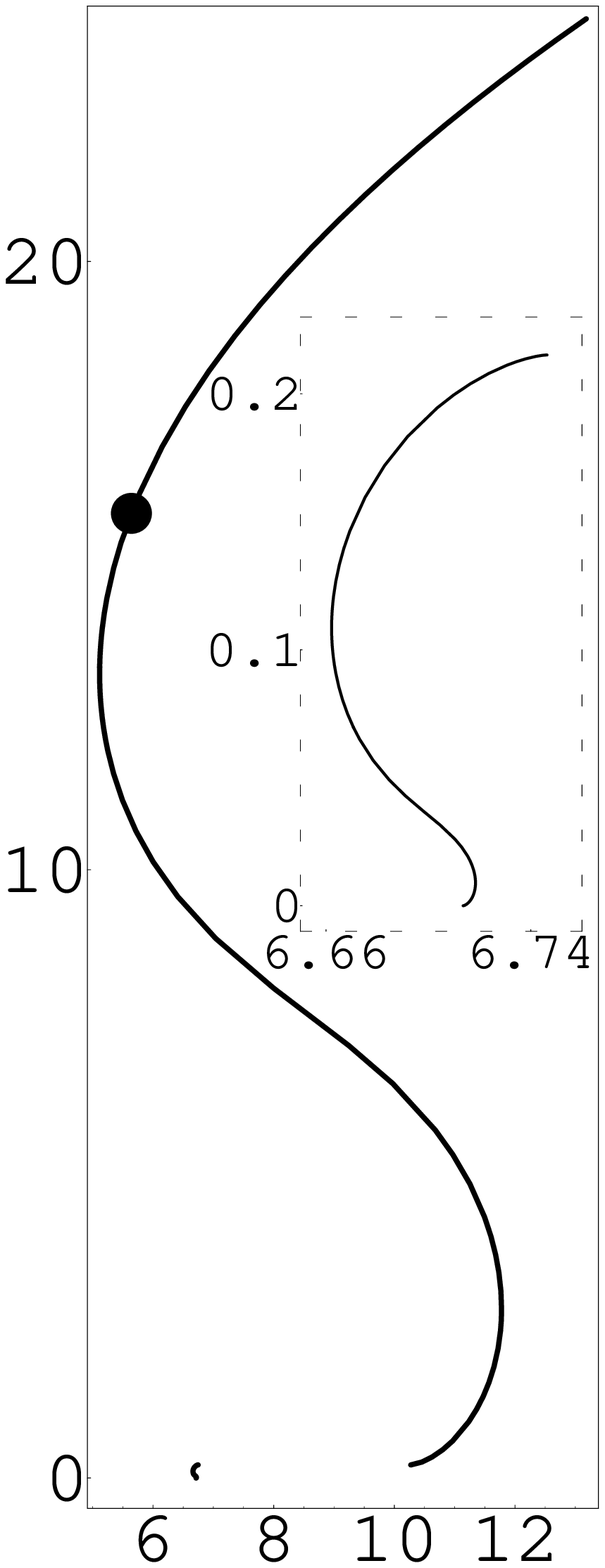}
\par {\small (e)}
\end{minipage}
\par
\begin{minipage}[b]{.224\hsize}
\centering\leavevmode \epsfxsize=\hsize \epsfbox{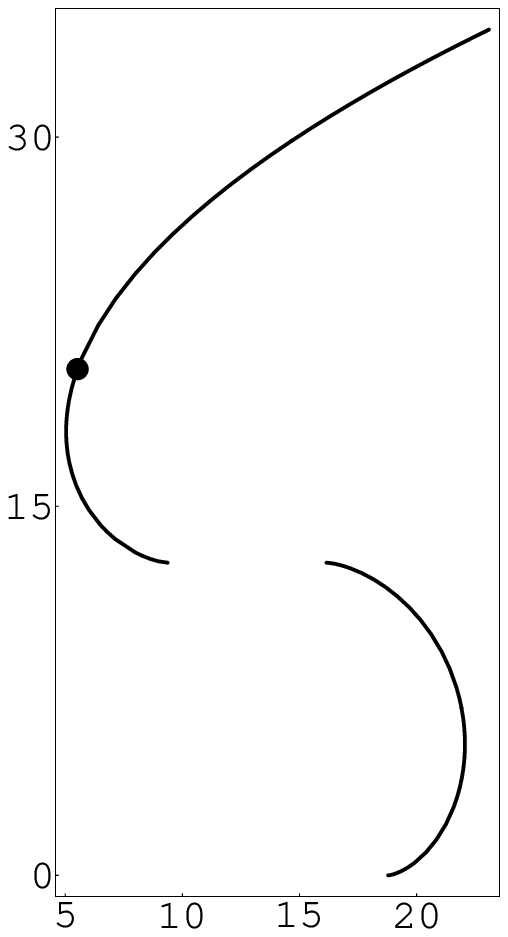}
\par {\small (f)}
\end{minipage}\hfill%
\begin{minipage}[b]{.119\hsize}
\centering\leavevmode \epsfxsize=\hsize \epsfbox{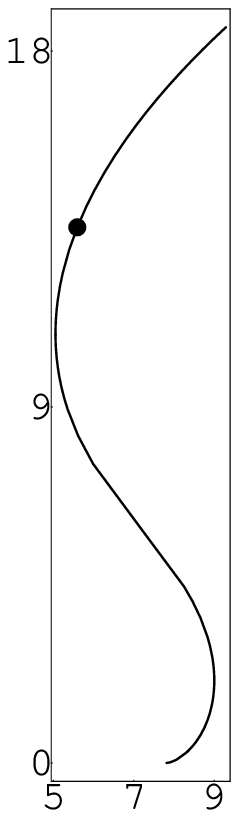}
\par {\small (g)}
\end{minipage}\hfill%
\begin{minipage}[b]{.152\hsize}
\centering\leavevmode \epsfxsize=\hsize \epsfbox{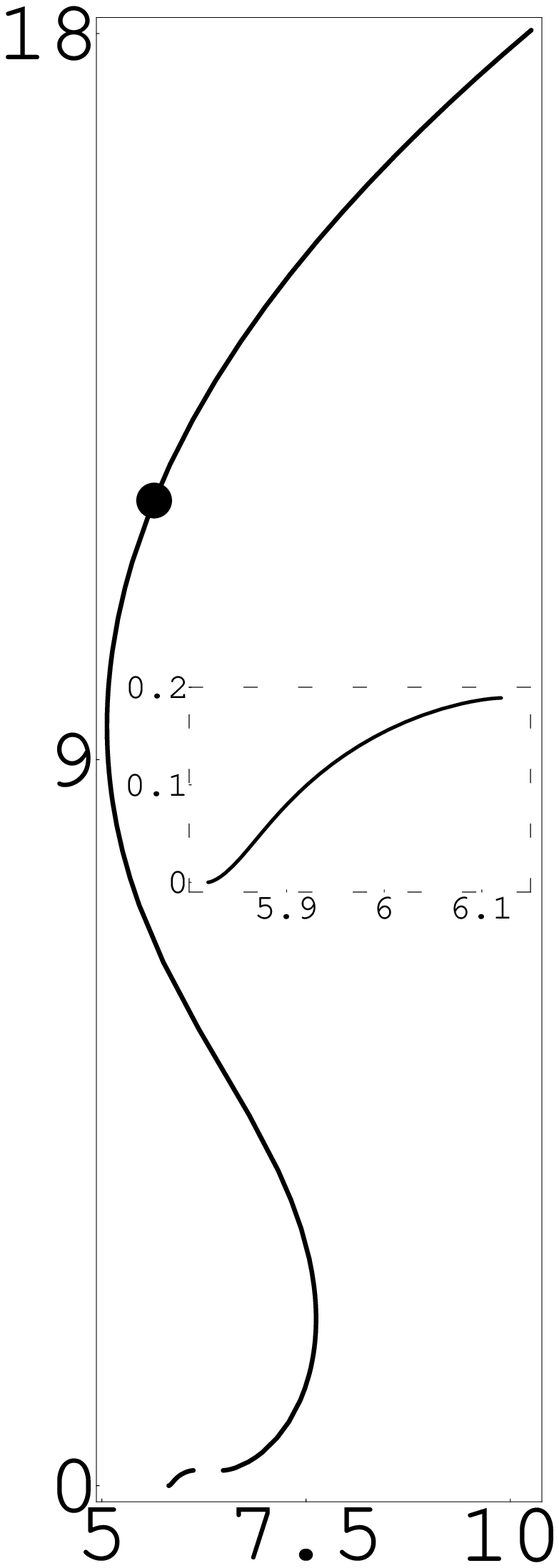}
\par {\small (h)}
\end{minipage}\hfill%
\begin{minipage}[b]{.1\hsize}
\centering\leavevmode \epsfxsize=\hsize \epsfbox{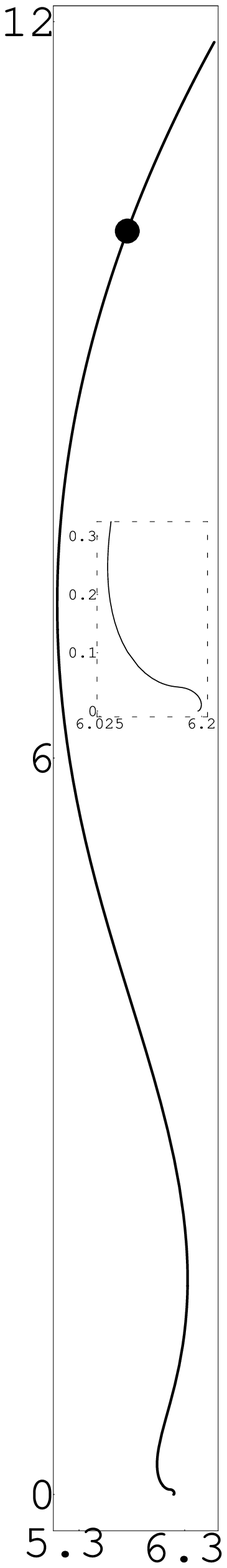}
\par {\small (i)}
\end{minipage}\hfill%
\begin{minipage}[b]{.36\hsize}
\centering\leavevmode \epsfxsize=\hsize \epsfbox{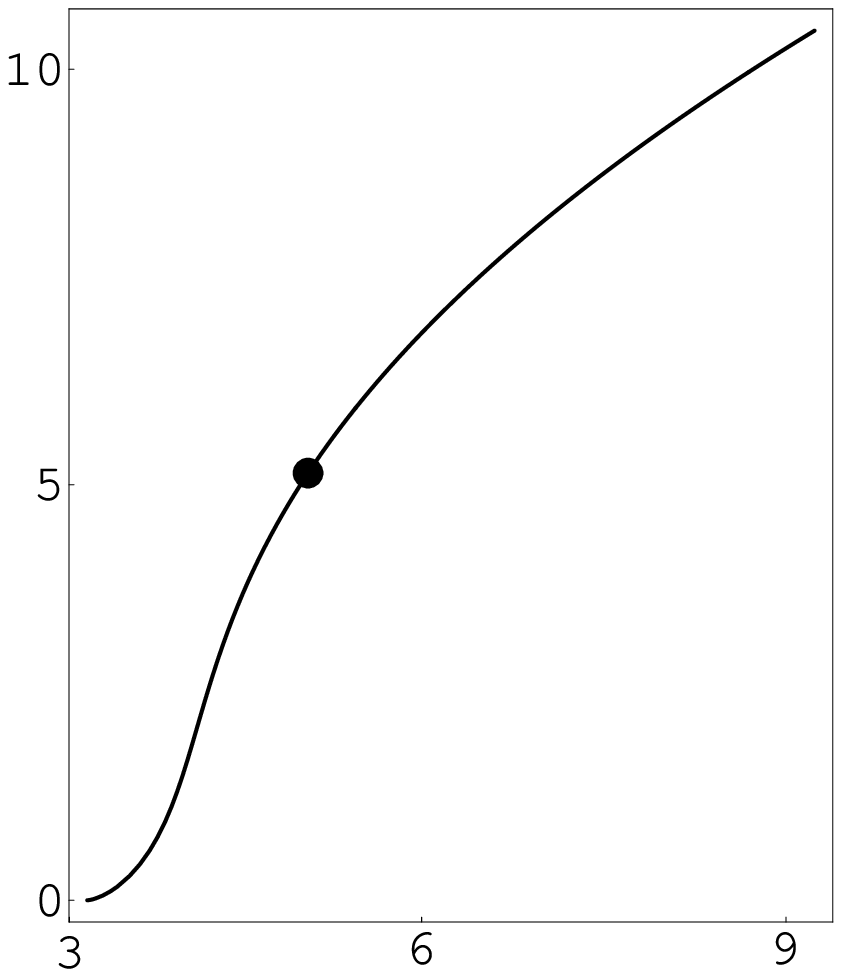}
\par {\small (j)}
\end{minipage}
\caption{Embedding diagrams of the types
  (a)~NS$_2$    ($a^2=1$,     $e^2=0.16$),
  (b)~NS$_3$    ($a^2=1.05$,  $e^2=0.16$),
  (c)~NS$_4$    ($a^2=0.82$,  $e^2=0.19$),
  (d)~NS$_5$    ($a^2=0.847$, $e^2=0.19$),
  (e)~NS$_6$    ($a^2=0.9$,   $e^2=0.19$),
  (f)~NS$_7$    ($a^2=0.77$,  $e^2=0.25$),
  (g)~NS$_8$    ($a^2=0.9$,   $e^2=0.25$),
  (h)~NS$_9$    ($a^2=0.99$,  $e^2=0.25$),
  (i)~NS$_{10}$ ($a^2=1.23$,  $e^2=0.25$), and
  (j)~NS$_{11}$ ($a^2=1$,     $e^2=0.9$).}
\label{ed-ns2-11}
\end{figure}

All    of    the     black-hole    diagrams    are    determined    by
Fig.\,\ref{r-a2-016}. In  order to cover all  of the naked-singularity
diagrams all Figs~\ref{r-a2-016}--\ref{r-a2-090}  have to be used. The
typical   diagrams   are    represented   in   Figs~\ref{ed-bh1}   and
\ref{ed-bh2-4} for the black-hole spacetimes, and in Figs~\ref{ed-ns1}
and \ref{ed-ns2-11}  for the  naked-singularity spacetimes. We  do not
explicitly consider the cases corresponding to situations when extrema
of  $a^2_{\rm e}(r;e)$  coincide in  an inflex  point as  they  can be
discussed in a straightforward manner.

\begin{figure}[t]
\begin{minipage}[c]{.4\hsize}
\centering \leavevmode \epsfxsize=\hsize \epsfbox{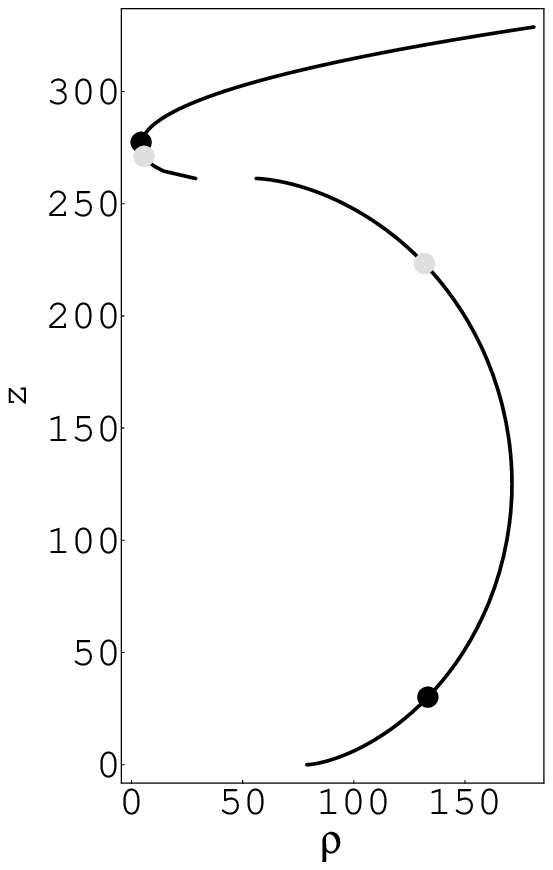}
\end{minipage}\hfill%
\begin{minipage}[c]{.57\hsize}
\centering \leavevmode \epsfxsize=.95\hsize \epsfbox{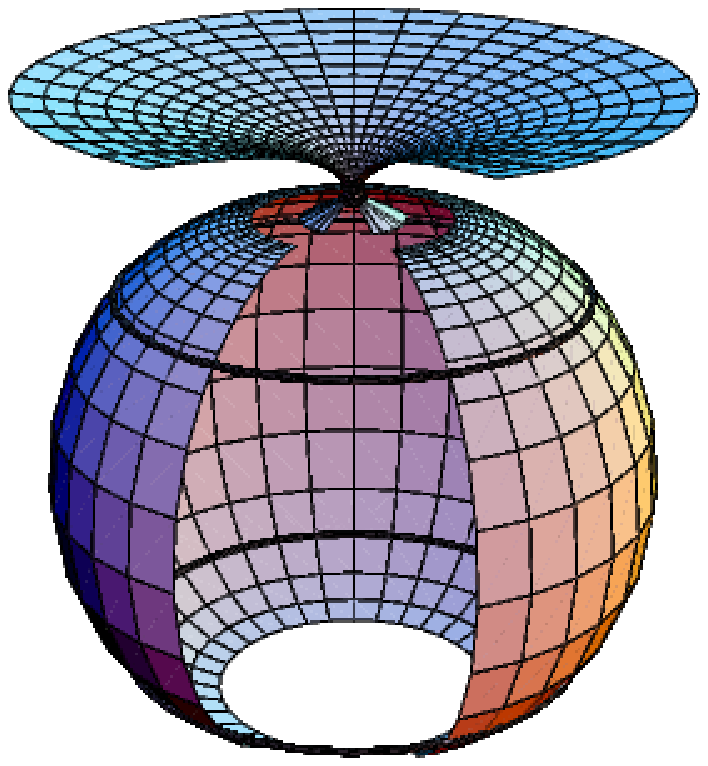}
\end{minipage}
\caption{Embedding diagrams  of   the  NS$_7$  type,  constructed  for
  $a^2=0.10005$, $e^2=0.9$.  The diagram is separated into two regions
  having one  throat and  one belly. All  four photon  circular orbits
  find  themselves in  the  embeddable region.   In  each region,  one
  corotating  orbit (gray  spots) and  one counterrotating  one (black
  spots) exist.}
\label{ed-ns7a}
\end{figure}

\begin{figure}[t]
\begin{minipage}[c]{.3\hsize}
\centering \leavevmode \epsfxsize=.9\hsize \epsfbox{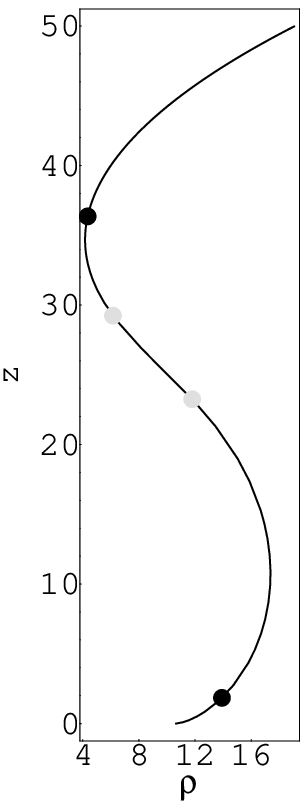}
\end{minipage}\hfill%
\begin{minipage}[c]{.67\hsize}
\centering \leavevmode \epsfxsize=.9\hsize \epsfbox{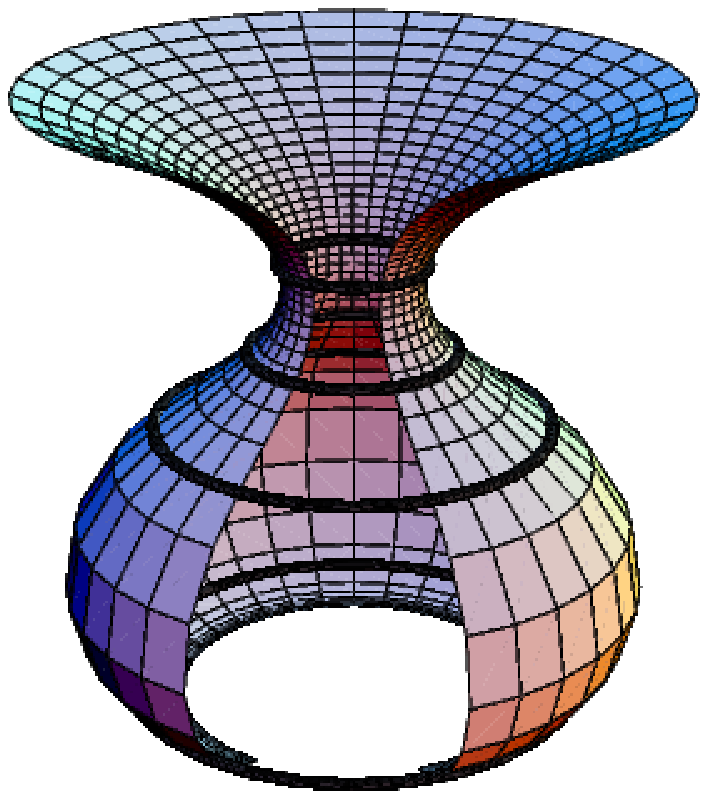}
\end{minipage}
\caption{Embedding diagrams  of   the  NS$_8$  type,  constructed  for
  $a^2=0.105$, $e^2=0.9$.  The diagram  is simply connected having one
  throat and  belly.  All four photon circular  orbits find themselves
  in  the  embeddable  region.   The  corotating  and  counterrotating
  circular  photon  orbits are  displayed  as  gray  and black  spots,
  respectively.}
\label{ed-ns8a}
\end{figure}

The  embeddability of  photon  circular orbits  can be  systematically
discussed by using Figs~\ref{r-a2cpe} and \ref{a2-e2}, and behaviour of
$a^2_{\rm e}(r;e)$  and $a^2_{\rm ph}(r;e)$ in  related situations. We
shall drop  off the  discussion here, we  only explicitly  mention two
interesting  cases  of  naked-singularity spacetimes  containing  four
photon  circular orbits  all of  which are  located in  the embeddable
parts of  the optical  geometry instead. (Note  that such  a situation
cannot occur  in Kerr backgrounds.) In the  first case ($a^2=0.10005$,
$e^2=0.9$),  there  are  two  separated  parts of  the  diagram,  both
containing two  of the circular orbits  (see Fig.\,\ref{ed-ns7a}), in
the second case ($a^2=0.105$, $e^2=0.9$),  there is only one region of
embeddability   containing   all   of   the   circular   orbits   (see
Fig.\,\ref{ed-ns8a}).   In  both   cases,   the  embedding   diagrams
necessarily have a throat and a belly.

\subsection{The classification}

The following classification will  be made according to the properties
of  the  embedding  diagrams.    The  Kerr\nd  Newman  spacetimes  are
characterised  by the  number  of embeddable  regions  of the  optical
geometry, and by the number  of turning points at these regions, given
successively for the regions  with descending radial coordinate of the
geometry  (which  are   presented  in  parentheses).   Therefore,  the
classification can be given in the following way:

\begin{center}
\begin{tabular}{llll}
\hline
BHs &  NSs &  NSs &  NSs                                \\ \hline
BH$_1$:  $1(1)$  &
       NS$_1$: $4(1,1,1,1)$ &
              NS$_5$:  $3(1,1,2)$ &
                     NS$_9$:\hphantom{${}_0$} $2(2,0)$  \\
BH$_2$:  $2(1,0)$ &
       NS$_2$:  $3(2,1,1)$  &
              NS$_6$: $2(2,2)$  &
                     NS$_{10}$:  $1(4)$                 \\
BH$_3$:  $2(1,2)$ &
       NS$_3$: $2(3,1)$  &
              NS$_7$: $2(1,1)$ &
                     NS$_{11}$: $1(0)$                  \\
BH$_4$: $3(1,1,1)$ &
        NS$_4$: $3(1,1,0)$ &
              NS$_8$: $1(2)$ &
                      {}                                \\ \hline
\end{tabular}
\end{center}

\begin{figure}[p]
\centering \leavevmode \epsfxsize=.95\hsize \epsfbox{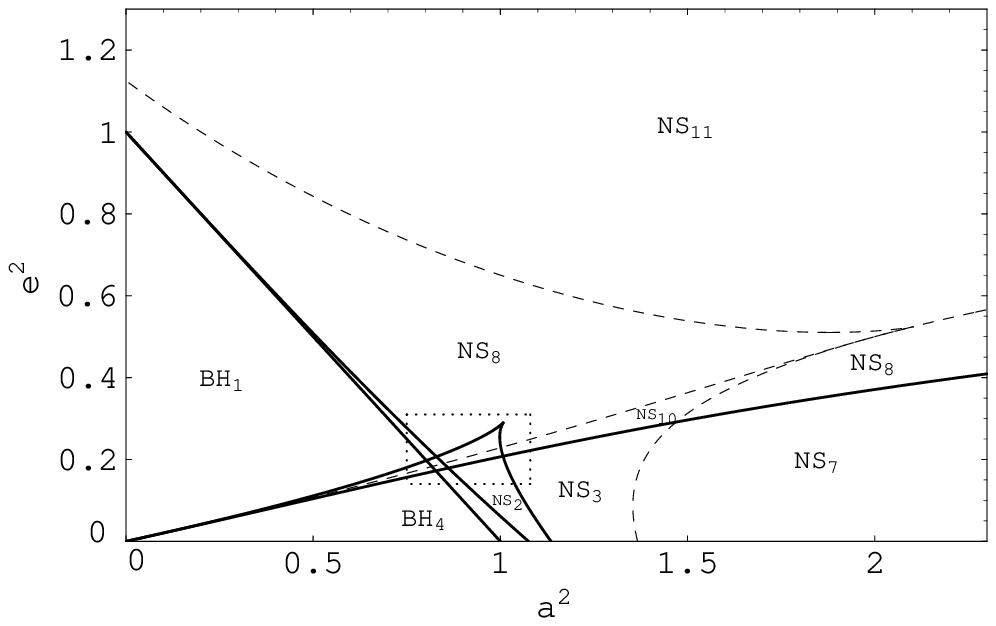}
\par\vskip 6mm\par
\centering \leavevmode \epsfxsize=.95\hsize \epsfbox{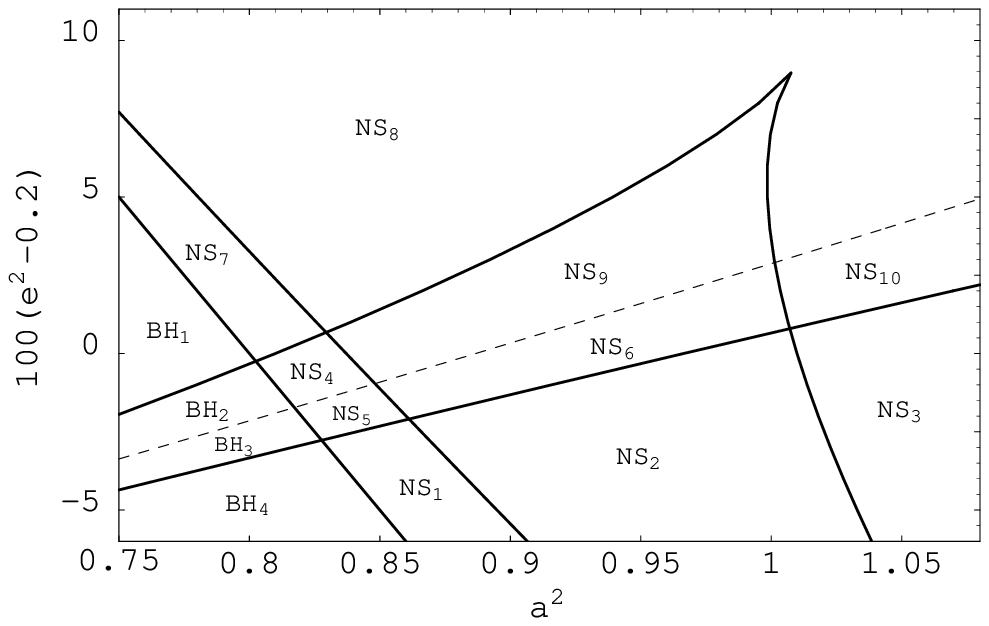}
\caption{Classification of the Kerr\nd Newman  spacetimes according to
  prop\-er\-ties of the embedding diagrams.  The solid lines separate
  spacetimes with different numbers of embeddable regions, the dashed lines
  separate spacetimes with different numbers of turning points (throats and
  bellies, in other words), at which the centrifugal force vanishes.  The
  parameter plane $a^2$-$e^2$ is divided into regions corresponding to types
  BH$_1$--BH$_4$ and NS$_1$--NS$_{11}$.  The central part marked by dotted
  rectangle is zoomed in the lower graph.}
\label{a2-e2cls}
\end{figure}

The parameter space of the  Kerr\nd Newman spacetimes can be separated
into  regions  corresponding  to  the classes  BH$_1$\nd  BH$_4$,  and
NS$_1$\nd NS$_{11}$ by a  numerical code expressing the extreme points
of  the  functions  $a^2_{\rm   e}(r;e)$  and  $a^2_{\rm  c}(r;e)$  as
functions of  $e^2$. Results  of the numerical  code are  presented in
Fig.\,\ref{a2-e2cls}, which represents the classification completely.

\section{Concluding remarks}                        \label{concluding}

In  the  rotating black-hole  and  naked-singularity backgrounds,  the
embedding   diagrams  of  the   optical  reference   geometry  reflect
immediately just one important property of the inertial forces related
to a test-particle circular motion.  Namely, the turning points of the
embedding  diagrams occur just  at radii  where the  centrifugal force
vanishes  and reverses  sign,  independently on  the  velocity of  the
motion. However, contrary to  the spherically symmetric black-hole and
naked-singularity  backgrounds, in the  rotating backgrounds  radii of
the photon circular orbits do not coincide with the radii of vanishing
centrifugal  force, and,  moreover,  in a  variety  of Kerr\nd  Newman
backgrounds,  some of these  orbits even  are not  located in  the the
regions  of  the  optical   geometry  that  are  embeddable  into  the
3-dimensional Euclidean space.

The properties of the embedding diagrams have been discussed, and a
clas\-si\-fica\-tion scheme of the Kerr\nd Newman backgrounds reflecting the
number of embeddable regions, and the number of turning points of these
regions, was presented. The ring singularity in all of the rotating
backgrounds, and the horizons of the black-hole backgrounds, are always
located outside the embeddable regions.  Further, embeddability of regions
containing the photon circular orbit has been established.

The presence of  a nonzero charge parameter $e$  in the Kerr\nd Newman
backgrounds  enriches  significantly  the  variety  of  the  embedding
diagrams  in  comparison  with  the  case  of  pure  Kerr  backgrounds
\cite{SH99c}.

For the black-hole backgrounds, there is just one embeddable region outside
the outer horizon, containing just one throat. However, under the inner
horizon, there can be two, one, or no embeddable regions. They can have a
throat and a belly. Above the outer horizon, the outermost, counterrotating
photon circular orbit is always embeddable, while the inner, corotating one
is embeddable in backgrounds with sufficiently small specific angular
momentum, but non-embeddable in the other ones. On the other hand, the
photon circular orbits existing under the inner horizon are always located
in the non-embeddable regions.

For the naked-singularity backgrounds, the variety of possible embedding
dia\-grams is much more complex and can be directly read out from
Figs~\ref{ed-ns1}--\ref{ed-ns2-11}.  Let us point out the most important
phenomena. There can exist diagrams consisting from four separated regions,
or, oppositely, simply connected diagrams, having two throats and two bellies.
On the other hand, there are also diagrams having no turning point. Moreover,
in some naked-singularity backgrounds containing four photon circular orbits,
all of the orbits are located in the embeddable regions---such situation is
impossible in the black-hole backgrounds.

The embedding diagrams of the optical geometry give an important tool of
visualisation and clarifying of the dynamical behaviour of test particles
moving along equatorial circular orbits, by imagining that the motion is
constrained on the surface $z(\rho)$~\cite{KSA}. The shape of the surface
$z(\rho)$ is directly related to the centrifugal acceleration. Within the
rising portions of the embedding diagram, the centrifugal acceleration
points towards increasing values of $r$, and the dynamics of test particles
has essentially Newtonian character. However, in the descending portions of
the embedding diagrams, the centrifugal acceleration has a radically
non-Newtonian character, as it points towards decreasing values of $r$.
Such a kind of behaviour appears where the diagrams have a throat or a
belly. At the turning points of the diagram, the centrifugal acceleration
vanishes and changes its sign, i.e., $\d r/\d\rho = \d z/\d\rho = 0$.

We can understand this connection of the centrifugal force and the
embedding of the optical space in terms of the radius of gyration
representing rotational properties of rigid bodies. In Newtonian mechanics,
it is defined as the radius $\tilde{R}$ of the circular orbit on which a
point-like particle having the same mass $M$ and angular velocity $\Omega$
as a rigid body would have the same angular momentum $J$:
\be
  J = M\tilde{R}^2\Omega.
\ee
Defining a specific angular momentum $\ell = J/M$, we obtain
\be
  \tilde{R} = \sqrt{\frac{\ell}{\Omega}}.                   \label{rogsam}
\ee
For a point-like particle moving on $r = {\rm const}$, there is $\tilde{R}
= r$.

In general relativity, the radius of a circle can be given by two standard
ways---namely as the circumferential radius, and as the proper radial
distance. A third way can be given by generalising Eq.~(\ref{rogsam}) for a
point particle moving along $r = {\rm const}$; we use $\ell = L/E$, where
$L$ is the angular momentum of the particle, and $E$ is its energy. [In
stationary spacetimes, the angular velocity have to be related to the
family of locally non-rotating observers, cf. Eq.~(\ref{tilOmRr}).] In
Newtonian theory, these three definitions of radius issue identical
results, however, in general relativity they are distinct. The radius of
gyration $\tilde{R}$ is convenient for discussing the dynamical effects of
rotation; the direction of increase of $\tilde{R}$ defines {\it local
  outward direction\/} of these effects. The surfaces $\tilde{R} = {\rm
  const}$, called von~Zeipel cylinders, were proved to be a very useful
concept in the theory of rotating fluids in stationary, axially symmetric
spacetimes. In Newtonian theory, they are ordinary straight cylinders, but
their shape is deformed by general-relativistic effects, and their topology
may be non-cylindrical. There is a critical family of self-crossing
von~Zeipel surfaces~\cite{KJA}.

It is crucial that in the Kerr\nd Newman spacetimes
\be
  [\tilde{h}_{\phi\phi}(\theta=\pi/2)]^{1/2} = \rho = \tilde{R},
\ee
i.e., the embedding diagrams of the equatorial plane of the optical
geometry are expressed in terms of the radius of gyration. Since in the
equatorial plane the centrifugal acceleration has the radial component
\be
  {\cal Z}(r) = \tilde{R}^{-1}\partial_r \tilde{R},
\ee
the relation of the embedding diagrams, the radius of gyration, and the
centrifugal force is clear. Note that the turning points of the embedding
diagrams determine both the radii where the centrifugal force changes sign,
and the radii of cusps where the critical von~Zeipel surfaces are
self-crossing. Therefore, the embedding diagrams also reflect properties of
fluid rotating in Kerr\nd Newman spacetimes.

Notice that the black-hole backgrounds have an unified character above the
event horizon---there exists a throat of the embedding diagram indicating
change of sign of the centrifugal force nearby the event horizon. On the
other hand, the naked-singularity spacetimes give a wide variety of the
behaviour of the embeddings and centrifugal forces, ranging from the simple
`Newtonian' backgrounds with no change of sign of centrifugal force, to
very complicated backgrounds, where the sign is changed four times.

We established also the embeddability of circular photon geodesics. There
is a lot of case when these orbits are located in regions that are
non-embeddable.

Notice that because the photon circular geodesics are given by the
condition ${\cal Z}(r)-{\cal C}(r)=0$ for corotating orbits, and ${\cal
  Z}(r)+{\cal C}(r)=0$ for counterrotating orbits, the corotating
(counterrotating) orbits are located on the descending (rising) portion of
the embedding diagrams (assuming ${\cal C}(r)<0$), if they enter the
embeddable regions---see Figs~\ref{ed-ns7a} and \ref{ed-ns8a}.

Finally, it should be noted that in the case of the spherically symmetric
spacetimes (\Sch\ and Reissner\nd Nordstr\"om \cite{KSA,MTW,dFYF}, and
Schwarzschild\nd de~Sitter \cite{SH99b,S99}), it can be shown that the
limits of embeddability of the optical geometry coincide with the existence
limit of static configurations of uniform density. Unfortunately, we
haven't found any evident connection between the embeddability limits of
the diagrams and any other phenomenon in the Kerr\nd Newman spacetimes.
Search for such phenomenon remains a big challenge for future
investigation.

\section*{Acknowledgements}

This  work  has  been  partly  supported  by  the  GA\v{C}R  grant  No
202/99/0261,  the  grant  J10/98:  192400004, and  the  Committee  for
collaboration  of Czech Republic  with CERN\@.   It has  been finished
during  visits at  The  Abdus  Salam ICTP,  Trieste,  and CERN  Theory
Division, Geneva;  the authors would  like to thank  both institutions
for perfect hospitality.  The authors would also like to express their
gratitude to Prof.~Marek Abramowicz for stimulating discussions.

\end{document}